\documentclass[journal]{IEEEtran}
\IEEEoverridecommandlockouts
\usepackage{setspace}
\usepackage{cite}
\usepackage{amsmath,amssymb,amsfonts}
\usepackage{multirow}
\usepackage[linesnumbered,ruled]{algorithm2e}
\usepackage{array}
\usepackage{url}
\usepackage{verbatim}
\usepackage{stfloats}
\usepackage{graphicx}
\usepackage{subfigure}
\usepackage{float}
\usepackage{textcomp}
\usepackage{xcolor}
\usepackage{soul}
\usepackage{booktabs}
\usepackage{bm}
\usepackage{multirow}
\usepackage[colorlinks, linkcolor=black, anchorcolor=black, citecolor=black]{hyperref}
\def\BibTeX{{\rm B\kern-.05em{\sc i\kern-.025em b}\kern-.08em
		T\kern-.1667em\lower.7ex\hbox{E}\kern-.125emX}}
\usepackage{balance}
\definecolor{matlabyellow}{rgb}{0.9290,0.6940,0.1250}
\begin{document}
	\sethlcolor{yellow} 
	
	\title{Scene-Conditioned PINN-GNN for Multipath RF Maps: Cross-Scene Generation and In-Scene Completion}

	\author{
		\IEEEauthorblockN{Lizhou Liu, Xiaohui Chen, Zihan Tang, Mengyao Ma, and Wenyi Zhang,~\IEEEmembership{Senior Member,~IEEE}}

		\vspace{-0.6cm}


		\thanks{Preliminary results of this work have been presented in part at 2025 International Conference on Future Communications and Networks \cite{PINN+GNN}. Supported by the National Key Research and Development Program of China (2025YFF0514400) and the National Natural Science Foundation of China (62231022).}
		\thanks{Lizhou Liu, Xiaohui Chen, and Wenyi Zhang are with the Department of Electronic Engineering and Information Science, University of Science and Technology of China, Hefei, Anhui 230026, China (e-mail: liulizhou@mail.ustc.edu.cn; cxh@ustc.edu.cn; wenyizha@ustc.edu.cn). \textit{(Corresponding author: Wenyi Zhang)}}
		\thanks{Zihan Tang and Mengyao Ma are with the Wireless Technology Lab, Huawei, Shenzhen 518129, China (e-mail: tangzihan1@huawei.com; ma.mengyao@huawei.com).}
	}
	
	\maketitle

	\begin{abstract}
		Radio frequency (RF) maps provide a compact representation of multipath propagation characteristics and are fundamental to channel modeling, coverage analysis, and environment-aware wireless optimization. This paper proposes a unified RF map construction framework based on a physics-informed neural network (PINN) and a graph neural network (GNN), supporting both cross-scene generation and in-scene completion with 2D and 2.5D environmental representations. The PINN embeds electromagnetic propagation constraints to establish a physically consistent mapping from receiver locations to multipath parameters, including path gain, time of arrival, and angles, while the GNN enforces spatial consistency by modeling correlations among neighboring receivers. To comprehensively evaluate multipath reconstruction quality, we propose a peak-weighted dynamic time warping metric that jointly accounts for amplitude errors and peak delay misalignment in channel impulse responses. Extensive experiments demonstrate that the proposed method consistently outperforms image-based, diffusion-based, and interpolation baselines across both map-level and multipath-level metrics, achieving robust generalization and high-fidelity RF map construction under sparse observations.
	\end{abstract}
	
	\begin{IEEEkeywords}
		Artificial intelligence (AI), graph neural network (GNN), multipath propagation, peak-weighted dynamic time warping, physics-informed neural network (PINN), radio frequency (RF) map.
	\end{IEEEkeywords}

	\vspace{-0.3cm}
	\section{Introduction}
	
	\IEEEPARstart{T}{o} meet the requirements of 6G networks for immersive interaction, ultra-reliable low-latency communications (URLLC), and high-precision localization, prior knowledge of spatial radio frequency (RF) propagation is essential \cite{6G_1}. As systems evolve toward massive multiple-input multiple-output (MIMO) and ultra-wide bandwidths, conventional pilot-based active measurements incur substantial signaling overhead and latency \cite{Wang2024}. Furthermore, deploying mobile access nodes like autonomous aerial vehicles (AAVs) requires proactive trajectory planning \cite{Ma2024}, necessitating the prediction of regional RF distributions before communication occurs. Consequently, wireless networks are shifting from measurement-driven reactive paradigms to environment-aware architectures \cite{Sun2025}.
	As a key enabler of this transition, RF maps provide a spatial representation of wireless propagation characteristics. Specifically, multipath RF maps incorporate diverse parameters including path gain, time of arrival (ToA), and angle of arrival (AoA). These maps facilitate multi-dimensional channel state inference and intelligent resource allocation with reduced overhead, supporting applications like high-precision localization and integrated sensing and communication (ISAC) \cite{Zeng2024}. However, constructing high-fidelity multipath RF maps in urban environments featuring complex blockages and rich scattering remains fundamentally challenging.

	Existing RF map construction generally falls into two categories: physics-based models and data-driven interpolation. Empirical models like Okumura–Hata \cite{Okumura}, \cite{Hata} and 3GPP Urban Macro \cite{3GPP} primarily address large-scale path loss, failing to capture site-specific multipath details in complex geometries. While ray tracing (RT) explicitly models physical interactions like reflection and diffraction \cite{Rizk1997}, \cite{Sionna}, its prohibitive computational overhead limits large-scale or real-time applications. 
	Alternatively, interpolation methods like Kriging \cite{Kriging}, tensor completion \cite{Schaufele2019}, and matrix completion \cite{Chouvardas2016} exploit spatial correlation for computational efficiency but lack physical awareness. In non-line-of-sight (NLoS) urban environments, these methods struggle with strong non-stationarity and shadow boundaries, yielding physically inconsistent RF maps under sparse sampling.

	Recent artificial intelligence (AI) advances introduce new paradigms for RF map construction. RadioUNet \cite{UNet} leverages deep neural networks to efficiently learn path loss distributions from simulation data. Building on this, cascaded U-Nets \cite{WNet} integrate environmental and sparse RT data for refined spatial predictions, while graph neural networks (GNNs) \cite{Chen2023} capture spatial topologies to enhance channel modeling. With the rapid development of generative models, RF map construction is increasingly formulated as a data generation problem. Generative adversarial networks (GANs) \cite{RME_GAN}, \cite{Chen2024}, \cite{Sarkar2024} and diffusion models \cite{RadioDiff}, \cite{Wang2025} are widely adopted for high-fidelity spatial generation and channel knowledge map (CKM) construction \cite{diffusion}, \cite{Zhao2026}.
	Despite these advances, most methods frame RF map construction as a computer vision (CV) image-to-image translation problem \cite{Wu2025}. These pixel-grid representations force continuous, floating-point physical measurements into quantized pixel intensities, inevitably losing fine-grained physical information during this image-centric transformation. Furthermore, they typically predict a single macroscopic quantity like path loss or received signal strength. While a few attempts incorporate spatial features like AoA \cite{Wang2025}, \cite{DoD}, they largely focus on the dominant path and overlook the multipath effects critical for practical wideband propagation \cite{use1}. Therefore, overcoming these image-centric quantization errors to construct high-fidelity, multi-parameter RF maps encompassing path gain, ToA, and AoA is essential for accurate wireless channel modeling. 


	Practical multipath RF map construction involves two challenging tasks. First, cross-scene generation requires predicting the global multipath distribution in uncharted areas using only accessible two-dimensional (2D) or three-dimensional (3D) environmental geometry to support proactive network deployment. Second, in-scene completion demands reconstructing high-resolution global channel distributions using strictly sparse measurements from mobile nodes when prior environmental information is unavailable.
	Under these data-restricted conditions, purely data-driven pixel-grid methods exhibit significant limitations. Lacking physical constraints and the ability to model non-Euclidean spatial topologies, they struggle to infer highly coupled multipath characteristics. Therefore, these approaches often yield estimation errors and physically inconsistent predictions that violate fundamental electromagnetic laws.

	Accurately characterizing complex multipath features requires integrating electromagnetic propagation mechanisms with spatial topology modeling. Physics-informed neural networks (PINNs) address this by embedding physical constraints into the learning process. Widely recognized in fluid dynamics \cite{PINN1}, PINNs are increasingly explored for wireless propagation \cite{PINN2}. By incorporating electromagnetic principles, like the geometric relationship among propagation distance, the speed of light, and ToA, as soft constraints, PINNs maintain physical consistency during data fitting and improve inference reliability.
	Furthermore, GNNs \cite{GNN1} expertly handle non-Euclidean topologies and model long-range node dependencies, effectively capturing spatial correlations induced by scattering and diffraction. Combining the physical constraint mechanisms of PINNs with the spatial reasoning capabilities of GNNs establishes a unified multipath RF map modeling framework. This integrated approach systematically overcomes the limitations of existing methods in complex urban environments under severely limited data conditions.

	To overcome the limitations of strictly pixel-grid representations and single-path predictions, we propose a unified PINN-GNN framework for RF map construction. This approach extends conventional modeling to joint multipath, multi-parameter characterization, simultaneously predicting path gain, ToA, and AoA. By directly utilizing original measurements and environmental semantics, the method extracts 2D or 2.5D geometric priors, including explicit height, from complex urban scenarios to assist LoS and NLoS identification. To decouple multipath parameters and prevent physical inconsistencies, PINNs embed electromagnetic propagation laws as soft constraints during data fitting. Subsequently, GNNs construct spatial graphs over sampling points and environmental features to model node correlations, ensuring the spatial consistency of the inferred parameters.
	We comprehensively evaluate this dual-consistency framework on cross-scene generation and in-scene completion tasks. Notably, to accurately characterize the temporal consistency of multipath structures, we propose a novel peak-weighted dynamic time warping (PW-DTW) metric. For cross-scene generation, comparing 2D and 2.5D maps verifies the critical role of explicit height information in recovering spatial multipath structures. Experimental results demonstrate that the proposed framework outperforms state-of-the-art learning and traditional interpolation methods in first-path prediction accuracy. Moreover, it achieves the stable, robust, and physically consistent inference of high-order multipath components in complex environments. Consequently, this work provides an effective solution for high-precision environment-aware communications and sensing in future 6G networks.

	The remaining part of this paper is organized as follows. Section \ref{sec:system_model} introduces the system model and formulates the multipath, multi-parameter RF map construction problem. Section \ref{section3} details the proposed scene-conditioned PINN-GNN framework. Section \ref{section4} outlines the baseline methods and evaluation metrics, while Section \ref{section5} presents experimental results and analysis. Finally, Section \ref{section6} concludes the paper.

	\vspace{-0.2cm}

	\section{System Model and Problem Formulation}
	\label{sec:system_model}

	This section establishes a unified system model for multipath RF map construction based on two environment representations for scene conditioning. Grounded in wireless propagation theory, the multipath channel is parameterized by channel gain, delay, and angular information. Subsequently, RF map completion and generation are formulated within a common framework, which aims to map scene conditions to multipath parameters under varying observation settings.

	\vspace{-0.4cm}

	\subsection{System Scenario and Environment Representation}
	We consider a wireless communication system consisting of a fixed transmitter (Tx) located at $\mathbf{t} \in \mathbb{R}^3$ and multiple receivers (Rx) distributed within a region of interest (RoI). The position of an arbitrary receiver is denoted by $\mathbf{r} \in \mathbb{R}^3$. To unify the description of signal propagation, the environmental geometry is modeled as a grid map, where each grid coordinate $(u,v)$ corresponds to a physical coordinate $(x,y)$. Depending on the available information granularity, the environment $\mathcal{E}$ can be represented in two forms:
	\begin{itemize}
		\item \textbf{2D environment map ($\mathcal{E}^{2\mathrm{D}}$)}: a binary occupancy map indicating buildings or accessible regions, which captures the topological structure of building boundaries and outdoor open spaces.
		\item \textbf{2.5D environment map ($\mathcal{E}^{2.5\mathrm{D}}$)}: an extension of the 2D map that additionally records the building height at each grid location. This representation explicitly characterizes variations in blockage severity and provides more reliable geometric cues for LoS visibility and dominant path formation, while avoiding the high cost and complexity of full 3D modeling.
	\end{itemize}

	To exclude non-target regions such as building interiors from prediction and evaluation, an effective region mask $R(u,v)\in\{0,1\}$ is defined for each scene, where $R(u,v)=1$ indicates a valid outdoor location. RF maps are constructed exclusively over the effective region defined by $R$.

	\vspace{-0.4cm}

	\subsection{Multipath Signal Model and Tensor Representation}
	In wireless signal propagation, information is encoded in both amplitude and phase of electromagnetic waves. Governed by Maxwell's equations, transmitted signals reach receivers via LoS and NLoS paths induced by reflection, scattering, and diffraction. When waves interact with environmental scatterers having different dielectric properties, they produce multipath components with varying delay, direction, and fading characteristics. The superposition of these multipath components in both temporal and spatial domains creates the received signal, a phenomenon formally known as multipath propagation.

	Consider a wideband wireless communication system where a Tx sends a baseband equivalent signal $s(t)$. For a Rx equipped with an $M \times N$ uniform planar array (UPA), the received signal $\mathbf{r}(t)$ under multipath propagation is given by
	\begin{equation}
		\vspace{-0.1cm}
		\mbox{\small$\displaystyle
		\mathbf{r}(t) = \sum_{l=1}^{L} \alpha_l \mathbf{a}(\theta_l, \varphi_l) s(t - \tau_l) + \mathbf{n}(t),
		\label{r(t)}
		$}
		\vspace{-0.1cm}
	\end{equation}
	where $L$ denotes the number of multipath components, $\alpha_l = |\alpha_l| e^{-j2\pi f_c \tau_l}$ is the complex gain of the $l$-th path with carrier frequency $f_c$ and delay $\tau_l$, $\theta_l \in [0,\pi]$ and $\varphi_l \in [-\pi, \pi]$ represent the elevation and azimuth angles, respectively, and $\mathbf{n}(t)$ is the additive white Gaussian noise (AWGN) vector. 

	For a UPA placed on the $x$--$y$ plane with antenna spacings $d_x$ and $d_y$, the array response vector $\mathbf{a}(\theta_l, \varphi_l) \in \mathbb{C}^{MN \times 1}$ is
	\begin{equation}
		\mbox{\small$\displaystyle
		\mathbf{a}(\theta_l, \varphi_l) =
		\left[e^{-j \frac{2\pi}{\lambda} (m d_x \sin \theta_l \cos \varphi_l + n d_y \sin \theta_l \sin \varphi_l)}\right]_{MN \times 1},
		\label{a}
		$}
	\end{equation}
	where $\lambda = c/f_c$ is the wavelength, $c$ is the speed of light, and $m \in \{0,1,\dots,M-1\}$ and $n \in \{0,1,\dots,N-1\}$ index the antennas along the $x$- and $y$-axes, respectively.

	The corresponding time-domain channel impulse response (CIR) is expressed as
	\begin{equation}
		\vspace{-0.1cm}
		\mbox{\small$\displaystyle
		\mathbf{h}(t)=\sum_{l=1}^{L} |\alpha_l| e^{-j2\pi f_c \tau_l} \mathbf{a}(\theta_l, \varphi_l) \delta(t - \tau_l),
		\label{h(t)}
		$}
		\vspace{-0.1cm}
	\end{equation}
	where $\delta(t)$ denotes the Dirac delta function. 
	
	As seen from \eqref{r(t)}--\eqref{h(t)}, the multipath channel is fully characterized by the physical parameters $\{\alpha_l, \tau_l, \theta_l, \varphi_l\}$.
	To facilitate neural network processing while preserving physical interpretability, the complex-valued channel is represented as a fixed-size real-valued tensor. For each receiver location $\mathbf{r}$, we define a multipath parameter tensor $\mathbf{Y}(\mathbf{r}) \in \mathbb{R}^{L \times 4}$ as
	\begin{equation}
		\mbox{\small$\displaystyle
		\mathbf{Y}(\mathbf{r}) = \left[ \mathbf{p}_1, \mathbf{p}_2, \dots, \mathbf{p}_L \right]^T,
		\label{eq:tensor_y}
		$}
	\end{equation}
	where each row $\mathbf{p}_l = [g_l, \tau_l, \theta_l, \phi_l]$ contains four key physical parameters of the $l$-th path: the path gain ($g_l = |\alpha_l|$), propagation delay ($\tau_l$), and angles of arrival (AoA), including the elevation angle $\theta_l$ and azimuth angle $\phi_l$.

	To accommodate spatially varying path counts, the tensor size is fixed to $L$ (e.g., $L=5$). Zero-padding is applied for fewer paths, whereas excess paths are truncated to retain the $L$ components with the highest received power. This representation enables the model to jointly learn the energy distribution and spatial directionality of multipath propagation.

	\vspace{-0.2cm}

	\subsection{Unified Task Statement}
	This work aims to generate or complete the multipath RF map across the target region. To unify different levels of information availability, all prior knowledge or observations are collectively referred to as \emph{conditioning information}, and a common output space $\mathbf{Y}$ is adopted to describe the task.

	Depending on the available conditioning information, the problem encompasses two representative operating modes:
	\begin{itemize}
		\item \textbf{Geometry-conditioned cross-scene generation}: In this setting, the environment map $\mathcal{E}$, represented as either $\mathcal{E}_{2\mathrm{D}}$ or $\mathcal{E}_{2.5\mathrm{D}}$, and the Tx configuration $\mathbf{t}$ are available, with no observations provided ($\mathcal{S}=\varnothing$). The objective is to directly generate the complete tensor $\mathbf{Y}$ over the effective region. This approach enables transferable prediction in unseen environments by learning the geometry-to-propagation mapping from extensive training data.
		\item \textbf{Measurement-conditioned in-scene completion}: When the environment map is unavailable ($\mathcal{E}=\varnothing$), the model relies on a small set of observed locations $\mathcal{S}=\{(\mathbf{r}_k, \mathbf{Y}_k)\}_{k=1}^K$. The goal is to exploit these sparse observations alongside physical constraints to recover the spatial distribution $\mathbf{Y}$ across the entire effective region.
	\end{itemize}

	To effectively accomplish these tasks and ensure physical consistency across both operating modes, the design and implementation details of the proposed scene-conditioned PINN--GNN framework are presented in Section \ref{section3}.

	\vspace{-0.26cm}

	\section{A Unified Scene-Conditioned PINN-GNN Framework}
	\label{section3}

\begin{figure*}[htbp]
		\centering
		\includegraphics[width=1\textwidth]{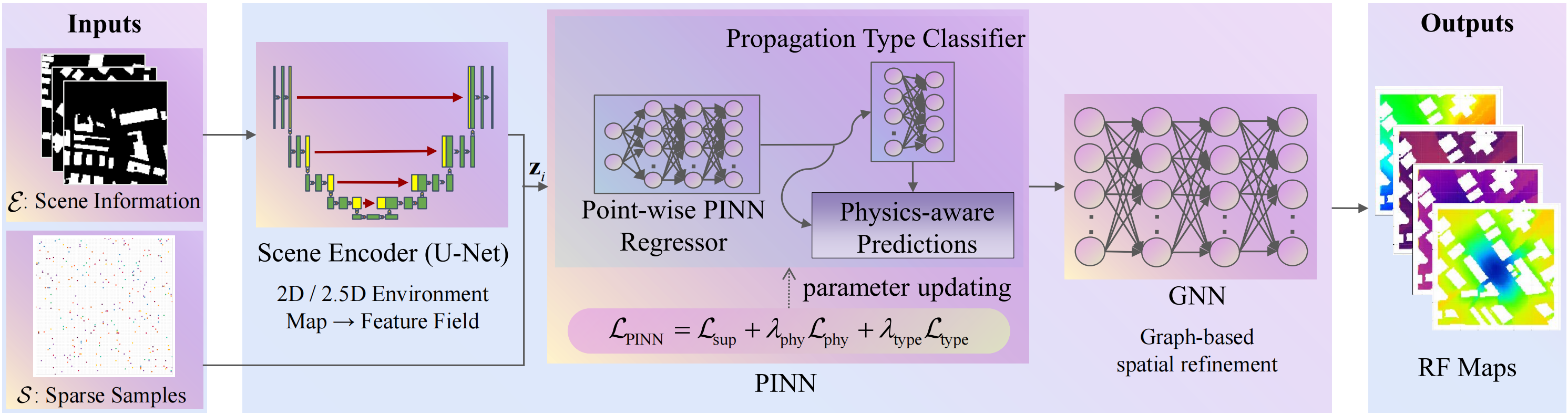}
		\vspace{-0.5cm}
		\caption{PINN-GNN architecture for RF map construction. Missing input modalities ($\mathcal{E}$ or $\mathcal{S}$) are automatically zero-padded to form the unified condition $\mathbf{z}_i$.}
		\label{fig:PINN_GNN}
		\vspace{-0.5cm}
	\end{figure*}

	This section presents a unified framework for constructing multipath RF maps. As illustrated in Fig. \ref{fig:PINN_GNN}, the cross-scene task takes the environmental information of the target scene as input, whereas the in-scene completion task relies on sparse sampled points within the scene and does not require explicit environmental information. The adaptive scene condition encoder serves as a unified input interface that flexibly processes heterogeneous inputs according to the current operating mode. Using the encoded conditions, a physics-guided predictor generates an initial estimate of the multipath parameter tensor by incorporating electromagnetic physical constraints. Finally, a graph-consistent refiner constructs a spatial graph over the predicted points and aggregates neighborhood information to suppress local noise and sharpen signal boundaries.

	\vspace{-0.4cm}

	\subsection{Overall Pipeline and Two Operating Modes}

	Let \(\mathcal{Q}_{\mathrm{valid}}\) denote the set of valid query locations determined by the environmental mask or validity rules. For any query location \(\mathbf{r}_i \in \mathcal{Q}_{\mathrm{valid}}\), the model outputs a predicted multipath tensor \(\hat{\mathbf{Y}}_i \in \mathbb{R}^{L \times Q}\), where $L$ denotes the number of paths and $Q$ is the dimension of the physical parameters for each path. The overall mapping process is formulated as
	\begin{equation}
		\mbox{\small$\displaystyle
		\label{eq:pipeline}
		\hat{\mathbf{Y}}_i = \mathcal{G}_{\Phi}\!\left(\tilde{\mathbf{Y}}_i,\ \mathcal{N}_k(i)\right), \quad
		\tilde{\mathbf{Y}}_i = \mathcal{F}_{\Theta}(\mathbf{r}_i, \mathbf{z}_i),
		$}
	\end{equation}
	where \(\mathcal{F}_{\Theta}\) denotes the point-wise predictor, \(\mathcal{G}_{\Phi}\) acts as a spatial refiner that aggregates local information from the \(k\)-nearest-neighbor set \(\mathcal{N}_k(i)\) of node \(i\), primarily responsible for mitigating local noise and enhancing spatial consistency. Specifically, \(\mathcal{F}_{\Theta}\) produces a physically plausible initial estimate \(\tilde{\mathbf{Y}}_i\) conditioned on the feature vector \(\mathbf{z}_i\). To provide a unified interface across different operating modes and govern model transferability, $\mathbf{z}_i$ is decomposed into geometric, environmental condition, and observation-guided features,
	\begin{equation}
		\mbox{\small$\displaystyle
		\mathbf{z}_i = [\mathbf{g}_i,\ \mathbf{e}_i,\ \mathbf{s}_i],
		$}
		\vspace{-0.2cm}
	\end{equation}
	where \(\mathbf{g}_i\) characterizes the Tx-Rx geometric relationship, supplying essential spatial cues and references for physical regularization. All spatial inputs are normalized to eliminate cross-scene scale discrepancies and enhance training stability. \(\mathbf{e}_i\) encodes the environmental geometry, and \(\mathbf{s}_i\) represents sparse observation guidance. To maintain a unified interface across operating modes, $\mathbf{e}_i$ and $\mathbf{s}_i$ automatically default to zero vectors when unavailable.

	Through this task-aware encoding mechanism, the framework switches functionality by adjusting the conditional features. When environmental maps are available but sparse observations are absent, the framework performs cross-scene generation, whereas given only sparse observations, the model performs in-scene completion.

	\vspace{-0.2cm}

	\subsection{Environment Encoding and Feature Alignment}
	\label{section3.2}


	To exploit environmental geometry and provide condition inputs with physical semantics for the subsequent PINN predictor, we design a unified U-Net-based environment encoder $\mathcal{E}_{\psi}$. This module transforms environmental maps into high-dimensional feature representations spatially aligned with query locations.

	In the cross-scene generation mode, the input is a 2D or 2.5D environmental grid tensor \(\mathcal{E} \in \mathbb{R}^{C_{\text{in}} \times H \times W}\), where \(H\) and \(W\) denote the height and width of the grid map, respectively, and \(C_{\text{in}}\) is the number of input channels, which adapts to the environment representation. Specifically, $C_{\text{in}}=1$ for a 2D binary occupancy mask, and $C_{\text{in}}=2$ for a 2.5D map combining a normalized height field and an occupancy mask. The encoder extracts multi-scale features via successive downsampling, while the decoder upsamples via transposed convolutions and fuses shallow geometric features through skip connections. This preserves global context and local boundaries, yielding a dense feature field with matched spatial resolution as
	\begin{equation}
		\mbox{\small$\displaystyle
		\mathbf{F} = \mathcal{E}_{\psi}(\mathcal{E}) \in \mathbb{R}^{D_f \times H \times W},
		$}
	\end{equation}
	where \(D_f\) denotes the dimensionality of the encoded feature channels. The resulting feature field \(\mathbf{F}\) maintains the same spatial resolution as the input environment, ensuring geometric alignment for subsequent feature sampling.

	Given small batch sizes in cross-scene training, convolutional blocks adopt group normalization rather than batch normalization, enhancing robustness against distribution shifts. To explicitly capture physical semantics like height and blockage, auxiliary supervision is applied during training, using lightweight prediction heads to reconstruct the input height field or occupancy mask.

	Since convolutions alone struggle with long-range LoS occlusions, we construct point-wise environmental features $\mathbf{e}_i$ via a ``local sampling + profile enhancement'' strategy. For an arbitrary query location \(\mathbf{r}_i\), local sampling is performed at the corresponding coordinates in the dense feature field \(\mathbf{F}\) using \(s \times s\) regional pooling, resulting in a location-aligned local feature vector \(\mathbf{f}_i \in \mathbb{R}^{D_f}\). To explicitly incorporate occlusions, we sample along the direct path between the Tx and Rx to form a profile feature $\mathbf{p}_i$, encoding statistics like the first obstruction, maximum blockage depth, and 2.5D clearance. The final vector is concatenated as $\mathbf{e}_i = [\mathbf{f}_i, \mathbf{p}_i]$.

	For in-scene completion where environmental maps are unavailable, $\mathbf{e}_i$ either defaults to $\mathbf{0}$ or is generated by a lightweight implicit encoder as an embedding $\mathbf{f}_{env,i} \in \mathbb{R}^{d_e}$, preserving the unified interface.

	\vspace{-0.26cm}

	\subsection{Physics-Guided Point-wise Predictor}

	The point-wise predictor, implemented via a PINN, serves as the core physical engine of the proposed framework. It is essentially a physics-guided multi-task neural network that aims to learn a nonlinear mapping from the unified scene-conditioned feature vector \(\mathbf{z}_i\) to the multipath parameter tensor \(\tilde{\mathbf{Y}}_i\), i.e.,
	\(\tilde{\mathbf{Y}}_i = \mathcal{F}_{\Theta}(\mathbf{r}_i, \mathbf{z}_i)\).
	By embedding electromagnetic propagation principles into both the network architecture and the loss design, this module is able to produce physically plausible inferences even in regions with sparse observations.

	\subsubsection{PINN Architecture}
	The point-wise predictor adopts a physics-guided multi-task multilayer perceptron (MLP) architecture to learn the mapping from point-level conditional features \(\mathbf{z}_i\) to the set of multipath parameters. The network input is the unified conditional feature vector \(\mathbf{z}_i\) constructed in Section \ref{section3.2}.

	\textbf{\textit{Point-wise PINN Regressor}:}   
	The available conditional features in $\mathbf{z}_i$ are first encoded and fused through attention to obtain a joint representation $\mathbf{h}_i$, which is then processed by a residual MLP to jointly regress the path gain, ToA, and angular parameters of all $L$ paths.

	\textbf{\textit{Propagation Type Classifier}:}
	In parallel, a lightweight MLP operating on the backbone features outputs \(\tilde{\mathbf{p}}_{i,l} \in \mathbb{R}^4\) for each path \(l\), representing the probabilities of LoS, reflection, scattering, and diffraction; these probabilities are used for auxiliary supervision and entropy-weighted physical regularization.

	\textbf{\textit{Physics-aware Predictions}:} 
	To ensure regressed outputs strictly conform to electromagnetic propagation characteristics, raw network outputs undergo dedicated activations and normalizations. A scaled hyperbolic tangent activation constrains path gain to a physically meaningful negative dynamic range consistent with path loss definitions. Similarly, a sigmoid function restricts ToA to a non-negative interval bounded by the maximum admissible delay. To prevent numerical instability from direct angular regression, the model predicts sine and cosine components, enforcing unit-circle normalization via projection. These transformations yield physically consistent multipath parameter estimates.

	\subsubsection{Training Objective and Loss Design}

	The network parameters are optimized by minimizing a composite loss function \(\mathcal{L}_{\text{PINN}}\). Let \(\mathcal{I}_{\text{sup}}\) denote the set of supervised samples, which depends on the operating mode. In the in-scene completion mode, \(\mathcal{I}_{\text{sup}}\) corresponds to the set of sparsely observed locations \(\mathcal{S}\); in the cross-scene generation mode, \(\mathcal{I}_{\text{sup}}\) consists of the sampled locations from the training scenes. The overall training objective is formulated as a weighted combination of a supervised loss term, a physics-based regularization term, and a propagation-type consistency term:
	\begin{equation}
		\mbox{$\displaystyle
		\mathcal{L}_{\text{PINN}} = \mathcal{L}_{\text{sup}} + \lambda_{\text{phy}}\mathcal{L}_{\text{phy}} + \lambda_{\text{type}}\mathcal{L}_{\text{type}}.
		$}
	\end{equation}

	\textbf{Supervised Loss (\(\mathcal{L}_{\text{sup}}\))}:  
	On the supervising set \(\mathcal{I}_{\text{sup}}\), the mean squared error (MSE) is adopted as
	\begin{equation}
		\mbox{$\displaystyle
		\mathcal{L}_{\text{sup}} = \frac{1}{|\mathcal{I}_{\text{sup}}|} \sum_{i\in\mathcal{I}_{\text{sup}}} \left\| \tilde{\mathbf{y}}_i - \mathbf{y}_i \right\|_2^2,
		$}
	\end{equation}
	where \(\tilde{\mathbf{y}}_i=\mathrm{vec}(\tilde{\mathbf{Y}}_i)\) and \(\mathbf{y}_i=\mathrm{vec}(\mathbf{Y}_i)\) denote the vectorized predicted and ground-truth multipath parameter tensors, respectively.

	\textbf{Propagation Type Consistency Loss (\(\mathcal{L}_{\text{type}}\))}:  
	For each location \(i\) and each path \(l\), the model outputs a probability vector over four propagation mechanisms,
	\begin{equation}
		\mbox{$\displaystyle
		\mathbf{p}_{i,l} = [p_{i,l}^{\text{LoS}},\, p_{i,l}^{\text{Refl}},\, p_{i,l}^{\text{Scat}},\, p_{i,l}^{\text{Diff}}], \quad \sum_k p_{i,l}^{k}=1.
		$}
	\end{equation}
	A cross-entropy loss is formulated as
	\begin{equation}
		\mbox{\small$\displaystyle
		\mathcal{L}_{\text{type}} = -\frac{1}{|\mathcal{I}_{\text{sup}}|} \sum_{i\in\mathcal{I}_{\text{sup}}}\sum_{l=1}^{L}\sum_{k=1}^{4} y_{i,l,k}\log\big(p_{i,l}^{k}+\epsilon\big),
		$}
	\end{equation}
	where \(y_{i,l,k}\) denotes weak labels constructed based on electromagnetic propagation rules, and \(\epsilon\) is a small constant introduced for numerical stability.

	\textbf{Physics Loss (\(\mathcal{L}_{\text{phy}}\)) }:
	To regularize predictions in unsupervised regions by enforcing fundamental electromagnetic propagation laws, the physics loss is introduced. 
	It consists of four complementary components, jointly formulated as
	\begin{equation}
		\mbox{$\displaystyle
		\mathcal{L}_{\text{phy}}
		=
		\lambda_1\mathcal{L}_{\text{CG}}
		+
		\lambda_2\mathcal{L}_{\text{ToA}}
		+
		\lambda_3\mathcal{L}_{\text{angle}}
		+
		\lambda_4\mathcal{L}_{\text{consist}}.
		$}
	\end{equation}
	Specifically, channel gain, ToA, and angular terms constrain the physical range and geometric consistency of individual paths, while the last consistency term enforces self-consistency across the multipath set. Propagation-type priors are incorporated via probability expectations, with constraint strength adaptively tuned using entropy-based confidence weighting.

	Since diverse propagation mechanisms like LoS, reflection, scattering, and diffraction exhibit fundamentally different attenuation and delay characteristics, uniform physical bounds cannot apply to all paths. For example, LoS components strictly follow geometric ToA constraints, whereas scattered paths experience significant delay spreading. To enable type-dependent physical regularization, we define reference coefficient vectors for each propagation mechanism as
	\begin{equation}
		\begin{aligned}
			\boldsymbol{\alpha} &= [\alpha_{\text{LoS}}, \alpha_{\text{Refl}}, \alpha_{\text{Scat}}, \alpha_{\text{Diff}}], \\
			\boldsymbol{\beta} &= [\beta_{\text{LoS}}, \beta_{\text{Refl}}, \beta_{\text{Scat}}, \beta_{\text{Diff}}].
		\end{aligned}
	\end{equation}
	Here, $\boldsymbol{\alpha}$ specifies the allowable upper-margin tolerance per path type relative to free-space channel gain, with tighter margins for LoS paths and larger deviations for diffracted components. Similarly, $\boldsymbol{\beta}$ defines lower-bound scaling factors relative to the geometric LoS ToA, where $\beta_{\text{LoS}}$ approaches unity and larger values apply to non-LoS components.

	The predicted propagation type \(\tilde{\mathbf{p}}_{i,l}\) is a continuous probability distribution rather than a discrete label. 
	To preserve differentiability during training, the discrete coefficients are transformed into continuous, path-level prior coefficients via probability-weighted expectations, which are expressed as
	\begin{equation}
		\bar\alpha_{i,l} = \sum_{k} p_{i,l}^{k}\alpha_{k}, \quad
		\bar\beta_{i,l} = \sum_{k} p_{i,l}^{k}\beta_{k},
	\end{equation}
	where \(k \in \{\text{LoS}, \text{Refl}, \text{Scat}, \text{Diff}\}\). 
	Through this formulation, the boundaries of the physical constraints are dynamically adjusted according to the model's confidence in the predicted propagation mechanism.

	To avoid imposing incorrect physical constraints during early training stages when the propagation-type predictions are highly uncertain, we introduce an entropy-based confidence gating mechanism. 
	The entropy of the predicted probability distribution and the corresponding adaptive confidence weight are defined as
	\begin{equation}
		\mbox{$\displaystyle
		H(\mathbf{p}_{i,l}) = -\sum_{k} p_{i,l}^{k}\log(p_{i,l}^{k} + \epsilon),
		$}
	\end{equation}
	\begin{equation}
		\mbox{$\displaystyle
		w_{i,l} = \left( 1 - \frac{H(\mathbf{p}_{i,l})}{\log 4} \right)^\eta,
		$}
	\end{equation}
	where \(\eta\) denotes a focusing factor. 
	When the predicted distribution approaches a uniform one, the entropy increases and the weight \(w_{i,l}\) vanishes, effectively disabling physical regularization and preventing erroneous guidance. 
	Conversely, when the predicted distribution becomes sharp, \(w_{i,l}\) approaches a zero-one degenerated distribution, and full-strength physical constraints are enforced.

	\textit{Out-of-Bound Penalties for Gain and ToA} (\(\mathcal{L}_{\text{CG}}, \mathcal{L}_{\text{ToA}}\)):
	Let the Tx–Rx distance be denoted by \(d_i\). 
	The theoretical LoS ToA is given by \(\tau_i^{\text{LoS}} = d_i / c\), where \(c\) is the speed of light. 
	The LoS reference power loss \(P_i^{\text{FSPL}}\) follows the free-space path loss model and is written as
	\begin{equation}
		\mbox{$\displaystyle
		P_i^{\text{FSPL}} = 20\log_{10}(d_i) + 20\log_{10}(f_c) + 20\log_{10}\!\left(\frac{4\pi}{c}\right),
		$}
	\end{equation}
	where \(f_c\) denotes the carrier frequency. 
	For each path \(l\), a smooth hinge penalty based on the softplus function \(\mathrm{softplus}(x)=\log(1+e^x)\) is adopted to penalize boundary violations. 
	Using the type-dependent coefficients \(\bar\alpha_{i,l}\) and \(\bar\beta_{i,l}\), the physical constraints on channel gain and ToA are formulated as \eqref{Lcg},
	\begin{figure*}[!hb]
		\vspace{-0.5cm}
		\hrulefill
		\begin{equation}
			\mbox{\small$\displaystyle
			\begin{aligned}
			\label{Lcg}
			\mathcal{L}_{\text{CG}}
			=
			\frac{1}{|\mathcal{I}_{\text{sup}}|}
			\sum_{i\in\mathcal{I}_{\text{sup}}}\sum_{l=1}^{L}
			w_{i,l}\,
			\mathrm{softplus}\!\left(
			\tilde{g}_{i,l} - \left(-P_i^{\text{FSPL}} + 10\log_{10}\bar\alpha_{i,l}\right)
			\right)^2.
			\end{aligned}
			$}
			\vspace{-0.5cm}
		\end{equation}
	\end{figure*}
	and
	\begin{equation}
		\mbox{$\displaystyle
		\mathcal{L}_{\text{ToA}}
		=
		\frac{1}{|\mathcal{I}_{\text{sup}}|}
		\sum_{i\in\mathcal{I}_{\text{sup}}}\sum_{l=1}^{L}
		w_{i,l}\,
		\mathrm{softplus}\!\left(
		\bar\beta_{i,l}\tau_i^{\text{LoS}} - \tilde{\tau}_{i,l}
		\right)^2,
		$}
	\end{equation}
	where \(\tilde{g}_{i,l}\) and \(\tilde{\tau}_{i,l}\) denote the predicted gain and ToA of the \(l\)-th path at location \(i\), respectively.

	\textit{Angular Geometric Consistency Loss} (\(\mathcal{L}_{\text{angle}}\)):  
	This term enforces geometric self-consistency between the predicted spatial arrival directions and the inferred propagation types. 
	A geometric consistency score \(s_i \in [0,1]\) is constructed based on the ToA and angular deviations of the first path relative to the geometric LoS reference, which is defined as
		\begin{equation}
			\mbox{\small$\displaystyle
			\begin{aligned}
			\label{si}
			s_i = \exp\!\left( -\frac{|\tilde{\tau}_{i,1} - \tau_i^{\text{LoS}}|}{\sigma_\tau} \right)
			\cdot
			\exp\!\left( -\frac{d_{\angle}(\tilde{\theta}_{i,1}, \tilde{\phi}_{i,1}; \theta_i^{\text{LoS}}, \phi_i^{\text{LoS}})}{\sigma_\angle} \right).
			\end{aligned}
			$}
		\end{equation}
	
	The angular distance metric accounting for phase periodicity is expressed as
		\begin{equation}
			\mbox{\small$\displaystyle
			\begin{aligned}
			\label{dangle}
			d_{\angle}
			=
			|\tilde{\theta}_{i,1} - \theta_i^{\text{LoS}}|
			+
			\min\!\left(
			|\tilde{\phi}_{i,1} - \phi_i^{\text{LoS}}|,
			360^\circ - |\tilde{\phi}_{i,1} - \phi_i^{\text{LoS}}|
			\right).
			\end{aligned}
			$}
		\end{equation}
	The score \(s_i\) intuitively measures the similarity between the predicted first path and an ideal LoS component.

	Using \(s_i\) as a soft label, a binary cross-entropy loss supervises the predicted LoS probability \(\tilde{p}_{i,1}^{\text{LoS}}\), which is written as \eqref{Langle}.
	\begin{figure*}[!hb]
		\begin{equation}
			\mbox{\small$\displaystyle
			\begin{aligned}
			\label{Langle}
			\mathcal{L}_{\text{angle}}
			=
			-\frac{1}{|\mathcal{I}_{\text{sup}}|}
			\sum_{i \in \mathcal{I}_{\text{sup}}}
			\Big[
			s_i \log(\tilde{p}_{i,1}^{\text{LoS}} + \epsilon)
			+
			(1 - s_i) \log(1 - \tilde{p}_{i,1}^{\text{LoS}} + \epsilon)
			\Big].
			\end{aligned}
			$}
			\vspace{-0.5cm}
		\end{equation}
	\end{figure*}
	This loss establishes closed-loop feedback where ToA or angular deviations reduces predicted LoS probabilities, while consistent estimates reinforce geometric alignment.

	\textit{Multipath Consistency Loss} (\(\mathcal{L}_{\text{consist}}\))  
	Beyond single-path constraints, the multipath parameter set should also satisfy structural consistency at the set level. 
	This term consists of a ranking consistency component and a gain–ToA coupling component, which are jointly defined as
	\begin{equation}
		\mathcal{L}_{\text{consist}} = \lambda_{\text{rank}}\mathcal{L}_{\text{rank}} + \lambda_{\text{pair}}\mathcal{L}_{\text{pair}}.
	\end{equation}

	To mitigate permutation ambiguity, paths are enforced into descending gain and ascending ToA orders. Slack variables $m_p$ and $m_\tau$ permit minor ordering violations, yielding the ranking loss as \eqref{Lrank}.
	\begin{figure*}[!hb]
		\begin{equation}
			\mbox{\small$\displaystyle
			\label{Lrank}
			\mathcal{L}_{\text{rank}}
			=
			\frac{1}{|\mathcal{I}_{\text{sup}}|}
			\sum_{i \in \mathcal{I}_{\text{sup}}}\sum_{l=1}^{L-1}
			\Big[
			\mathrm{softplus}(\tilde{g}_{i,l+1} - \tilde{g}_{i,l} + m_p)
			+
			\mathrm{softplus}(\tilde{\tau}_{i,l} - \tilde{\tau}_{i,l+1} + m_\tau)
			\Big].
			$}
		\end{equation}
		\begin{equation}
			\mbox{\small$\displaystyle
			\begin{aligned}
			\label{Lpair}
			\mathcal{L}_{\text{pair}}
			=
			\frac{1}{|\mathcal{I}_{\text{sup}}|}
			\sum_{i \in \mathcal{I}_{\text{sup}}}
			\frac{1}{|\mathcal{P}|}
			\sum_{(l,m) \in \mathcal{P}}
			\mathrm{softplus}\!\left(
			-\frac{(\tilde{g}_{i,l} - \tilde{g}_{i,m})(\tilde{\tau}_{i,m} - \tilde{\tau}_{i,l})}{s_p s_\tau}
			\right).
			\end{aligned}
			\vspace{-0.5cm}
			$}
			\vspace{-0.2cm}
		\end{equation}
	\end{figure*}

	Physically, longer ToA paths typically traverse greater distances or undergo more reflections, yielding higher attenuation. To penalize violations, we formulate a pairwise consistency loss \eqref{Lpair}, where $\mathcal{P} = \{(l,m)\mid 1 \le l < m \le L\}$ defines all path pairs, and training-set standard deviations $s_p$ and $s_\tau$ normalize scale differences.

	\vspace{-0.5cm}
	\subsection{Spatial Refinement via Graph Learning}

	Although the PINN predictor incorporates physical constraints, it fundamentally operates in a point-wise manner, treating each query location separately and neglecting the strong spatial correlation induced by the continuous variation of wireless channels. To address this limitation and further suppress local prediction noise, a GNN is introduced to refine the PINN outputs by enforcing spatial consistency.

	Specifically, a $k$-nearest neighbor graph $G = (V, E)$ is constructed over the valid query set $\mathcal{Q}_{\text{valid}}$ to explicitly model the topological relationship among receivers. Each node $v_i \in V$ corresponds to a valid receiver location, while edges in $E$ connect the $k$ closest neighbors according to the Euclidean distance in the coordinate space. In 2D scenarios, the distance metric is defined on the $(x, y)$ plane, whereas in 2.5D settings, the $(x, y, h)$ coordinates are used, where the height $h$ is normalized to the same scale as the planar coordinates to ensure isotropic neighborhood construction. The initial node features are primarily given by the point-level predictions of the PINN, denoted as $\tilde{\mathbf{y}}_i \in \mathbb{R}^{4L}$, optionally concatenated with a small set of geometric features to enhance discrimination in non-stationary regions such as building boundaries.

	The graph refiner employs four GraphSAGE \cite{Hamilton2017} layers with 128-dimensional hidden features over a (k=16) nearest-neighbor graph to aggregate neighborhood information. To preserve the physical priors embedded in the PINN estimates, it adopts residual learning and predicts only a correction term $\Delta \mathbf{y}_i$ rather than directly regressing the final multipath parameters. The refined output is thus expressed as
	\begin{equation}
		\hat{\mathbf{y}}_i = \tilde{\mathbf{y}}_i + \Delta \mathbf{y}_i,
	\end{equation}
	where $\hat{\mathbf{y}}_i$ denotes the final multipath parameter estimation. This residual formulation encourages the graph network to focus on learning spatially consistent adjustments, thereby reducing the risk of violating the physical plausibility of the point-wise predictions.

	To ensure training stability, the objective function of the refinement stage, denoted by $\mathcal{L}_{\text{graph}}$, is defined by minimizing the MSE between the refined predictions and the ground truth over the supervised sample set $\mathcal{I}_{\text{sup}}$, given as
	\begin{equation}
		\mbox{$\displaystyle
		\mathcal{L}_{\text{graph}} = \frac{1}{|\mathcal{I}_{\text{sup}}|} \sum_{i \in \mathcal{I}_{\text{sup}}} \|\hat{\mathbf{y}}_i - \mathbf{y}_i\|_2^2.
		$}
	\end{equation}

	\subsection{Joint Construction of RF Map via PINN and GNN}

	As illustrated in the system architecture, the proposed framework constructs high-fidelity RF maps by integrating physical consistency with spatial correlation.
    First, the framework processes the raw environment map \(\mathcal{E}\) and geometric coordinates into unified condition vectors \(\mathbf{z}_i\). A point-wise PINN then independently estimates the multipath parameters \(\tilde{\mathbf{Y}}_i\) for each location, enforcing physical validity via embedded constraints such as channel gain-ToA bounds and LoS alignment.
    To overcome the limitation of isolated point-wise predictions, a GNN is subsequently employed to refine these estimates. By constructing a \(k\)-nearest neighbor graph over the domain, the GNN models spatial dependencies and learns a residual correction \(\Delta \mathbf{Y}_i\), effectively smoothing out local noise while preserving the physical structure.
    

	During training, the scene encoder, PINN, and GNN are jointly optimized in an end-to-end manner. In addition to the main PINN-GNN objectives, the scene encoder is regularized by the auxiliary height/occupancy reconstruction supervision described in Section \ref{section3.2}, encouraging physically meaningful environmental features such as blockage and diffraction cues. The complete joint training and inference procedure is summarized in Algorithm \ref{alg:PINN_GNN_scene}.

	\begin{algorithm}[t]
		\caption{RF Map Construction with Scene-Conditioned Joint PINN--GNN}
		\small
		\label{alg:PINN_GNN_scene}
		\KwIn{
		Scene information $\mathcal{E}$ (2D / 2.5D environment map) \textbf{or} sparse samples $\{\mathbf{x}_m, \mathbf{y}_m\}_{m=1}^M$
		}
		\KwOut{
		RF map or point-wise predictions $\hat{\mathbf{Y}}$
		}

		\tcp{\textbf{Condition Initialization}}
		$\mathbf{e}_i \leftarrow$ Extract environment field $\mathbf{F} = \mathcal{E}_{\psi}(\mathcal{E})$ (if available, else $\mathbf{0}$)\;
		$\mathbf{s}_i \leftarrow$ Encode sparse measurements (if available, else $\mathbf{0}$)\;

		\tcp{\textbf{Query Construction}}
		Construct input queries $\{\mathbf{q}_i\}$ from either
		(i) grid locations implied by $\mathcal{E}$ or
		(ii) given sparse coordinates $\mathbf{x}_m$\;

		\tcp{\textbf{Joint PINN-GNN Training}}
		\For{each training epoch}{
			\textbf{Point-wise PINN inference:} \\
			Obtain physics-aware multipath predictions
			$\tilde{\mathbf{Y}}_i$;

			\textbf{Graph construction:} \\
			Build spatial neighborhood graph $G = (V, E)$
			according to query locations;

			\textbf{Graph-based refinement:} \\
			Refine point-wise predictions via message passing
			$\hat{\mathbf{Y}}$;

			\textbf{Loss computation:}\\
			Compute PINN loss $\mathcal{L}_{\text{PINN}}$ and GNN loss $\mathcal{L}_{\text{GNN}}$ \;

			\textbf{Parameter update:} \\
			Update all trainable parameters via backpropagation;
		}
		\tcp{\textbf{RF Map Generation / Completion}}
		Freeze model parameters\;
		Construct inference queries $\{\mathbf{q}_i^{\text{test}}\}$ over the target scene\;
		Predict $\tilde{\mathbf{Y}}_i$\;
		Refine via GNN to obtain $\hat{\mathbf{Y}}_i$\;
		Post-processing and inverse normalization\;
		\Return $\hat{\mathbf{Y}}$\;
	\end{algorithm}

	\section{Comparison Schemes and Evaluation Metrics}
	\label{section4}

	\subsection{Baseline Methods and Ablation Studies}

	Existing studies on RF map construction mainly focus on predicting the received power or path loss of the first-arriving path under cross-scene conditions, with the core objective of generating continuous radio coverage distributions given only environmental information. Since no prior work is capable of directly predicting the complete multipath parameter tensor in a cross-scene setting, a fair comparison with existing methods requires following their established evaluation protocol. Accordingly, the channel gain corresponding to the first path is extracted from the multipath predictions of the proposed framework, and the comparison is conducted at this scalar level.

	Under this alignment setting, RF map construction is formulated as a two-dimensional image generation or completion problem conditioned on environmental representations, where the goal is to predict first-path power maps or path loss maps. Three representative sampling-free, image-based methods are selected as cross-scene baselines.

	\begin{itemize}

		\item \textbf{RadioUNet \cite{UNet} (Supervised Regression):}
		RadioUNet is a classical end-to-end image regression approach that formulates RF map construction as a direct mapping from environmental layouts to path loss maps. It adopts a fully convolutional U-Net as the backbone and learns the nonlinear relationship between environmental structures and radio attenuation through pixel-wise supervision. Owing to its simplicity and effectiveness, RadioUNet has been widely used as a baseline in prior studies.

		\item \textbf{RME-GAN \cite{RME_GAN} (Adversarial Generation):}
		RME-GAN represents RF map generation methods based on conditional GANs. Through adversarial training, it is able to better capture high-frequency details in path loss distributions. The original RME-GAN relies on sparse path loss samples as conditional inputs. To ensure fairness in the sampling-free cross-scene setting, only environmental information is retained as the conditioning input in our experiments.

		\item \textbf{RadioDiff \cite{RadioDiff} (Diffusion-Based Generation):}  
		RadioDiff is a recent sampling-free RF map construction method based on denoising diffusion probabilistic models. It formulates RF map generation as a conditional diffusion process and progressively recovers high-fidelity first-path power maps through multi-step reverse denoising. Its architecture integrates attention mechanisms and frequency-domain modeling modules, achieving strong performance in both structural similarity and prediction accuracy.

	\end{itemize}

	It is worth noting that all the above cross-scene baseline methods are designed exclusively for modeling first-path power or path loss, and do not involve tensor-level prediction of multipath parameters. Therefore, they only participate in the aligned comparison at the first-path level.

	To ensure consistency and fairness across different methods, a unified alignment protocol is adopted for cross-scene first-path power evaluation. At the variable level, all comparisons are strictly restricted to the first-path power map $G$. For the proposed method, only the channel gain component $\hat{G}$ corresponding to the first path is extracted from the predicted multipath parameter tensor $\hat{\mathbf{Y}}$ for evaluation, ensuring consistency with methods that solely predict first-path power or path loss maps. At the spatial level, all image-based evaluation metrics are computed only within the receiver region $\Omega_R = \{(u,v)\mid R(u,v)=1\}$ to avoid interference from unreachable areas. At the normalization level, the structural similarity index (SSIM) is computed after min--max normalization within $\Omega_R$ for each scene, thereby mitigating the impact of dynamic range variations across different environments.

	To systematically evaluate each component of the proposed unified framework, we report results under identical cross-scene settings using 2D and 2.5D representations. Accordingly, three ablation schemes analyze the distinct impacts of physical and spatial consistency modeling.

	\begin{itemize}

		\item \textbf{w/o PINN:} The physical consistency constraint is removed, and the model relies solely on data-driven regression without incorporating physics-based loss terms. This setting is used to assess the impact of physical priors on the stability and generalization of multipath parameter learning.

		\item \textbf{w/o GNN:} The graph-based spatial refinement module is removed, and RF maps are generated directly from point-wise predictions. This configuration evaluates the role of explicit spatial relationship modeling in continuous map construction.

		\item \textbf{Separate:} A decoupled learning scheme is adopted as a baseline, where physics-aware regression and graph-based propagation are conducted in two independent stages. Specifically, a PINN is first trained to predict multipath parameters at each location under physical constraints, after which its outputs are fixed and a GNN is trained for neighborhood aggregation and spatial refinement. Unlike the proposed unified framework, this scheme does not employ end-to-end joint optimization and does not allow feature sharing across modules.

	\end{itemize}

	In addition to cross-scene generation, the proposed method is further evaluated under a sparse-sample completion setting. In this scenario, besides the above ablation baselines, the classical spatial interpolation method \textbf{Kriging} is included for comparison.

	\vspace{-0.5cm}
	\subsection{Evaluation Metrics}

	We adopt a two-level evaluation protocol to comprehensively assess the performance of the proposed method under different task settings. The first level consists of \textit{map-level metrics} for aligned cross-scenario comparison on first-path gain prediction, while the second level includes \textit{tensor-level metrics} for evaluating the reconstruction accuracy of complete multipath parameters. Furthermore, the proposed PW-DTW metric is utilized to evaluate multipath temporal consistency.

	\subsubsection{Map-Level Metrics (First-Path Gain)}

	Let $G(u,v)$ and $\hat{G}(u,v)$ denote the ground-truth and predicted first-path gain maps, respectively. All map-level metrics are computed only within the receiver region
	$\Omega_R = \{(u,v)\mid R(u,v)=1\}$.

	The root mean square error (RMSE) is defined as
	\begin{equation}
	\mbox{\small$\displaystyle
	\mathrm{RMSE}(G,\hat{G}) =
	\sqrt{\frac{1}{|\Omega_R|}\sum_{(u,v)\in\Omega_R}
	\left(\hat{G}(u,v)-G(u,v)\right)^2 }.
	$} 
	\end{equation}

	The normalized mean square error (NMSE) is given by
	\begin{equation}
	\mbox{\small$\displaystyle
	\mathrm{NMSE}(G,\hat{G}) =
	\frac{\sum_{(u,v)\in\Omega_R}
	\left(\hat{G}(u,v)-G(u,v)\right)^2}
	{\sum_{(u,v)\in\Omega_R} G(u,v)^2}.
	$}
	\end{equation}

	To measure structural similarity, the structural similarity index (SSIM) is employed. To mitigate the impact of dynamic range variations across different scenarios, both the ground-truth and predicted maps are first min--max normalized within $\Omega_R$ for each scenario as
	\vspace{-0.1cm}
	\begin{equation}
	\mbox{\small$\displaystyle
	G^{\text{norm}} = \frac{G - G_{\min}}{G_{\max} - G_{\min} + \epsilon},
	\hat{G}^{\text{norm}} = \frac{\hat{G} - \hat{G}_{\min}}{\hat{G}_{\max} - \hat{G}_{\min} + \epsilon}.
	$}
	\end{equation}
	Then, the SSIM metric is computed between the normalized maps $x = G^{\text{norm}}$ and $y = \hat{G}^{\text{norm}}$. The standard SSIM formulation is applied as
	\begin{equation}
	\mbox{\small$\displaystyle
	\mathrm{SSIM}(x,y)=
	\frac{(2\mu_x\mu_y+C_1)(2\sigma_{xy}+C_2)}
	{(\mu_x^2+\mu_y^2+C_1)(\sigma_x^2+\sigma_y^2+C_2)},
	$}
	\end{equation}
	where $\mu_x$ and $\mu_y$ denote the local means, $\sigma_x^2$ and $\sigma_y^2$ denote the local variances, and $\sigma_{xy}$ represents the local covariance of $x$ and $y$. These local statistics are computed using an $11 \times 11$ Gaussian sliding window. The constants are set to $C_1 = (0.01L_{\max})^2$ and $C_2 = (0.03L_{\max})^2$ with $L_{\max}$ being the dynamic range of the normalized images.

	\subsubsection{Tensor-Level Metrics (Multipath Parameters)}

	Let $\mathbf{Y}$, $\hat{\mathbf{Y}} \in \mathbb{R}^{H\times W\times L\times Q}$ denote the ground-truth and predicted multipath parameter tensors, respectively, where $L$ is the number of paths and $Q$ denotes the dimensionality of physical parameters per path. All tensor-level metrics are evaluated only within the effective propagation region $\Omega_M$, which is automatically inferred from the environmental map.

	Prior to calculating error metrics, all physical parameters are normalized using their respective statistical mean and standard deviation along each feature dimension to avoid bias caused by inconsistent units. The tensor-level RMSE is defined as
	\vspace{-0.1cm}
	\begin{equation}
	\mbox{\small$\displaystyle
	\mathrm{RMSE}_{\text{ten}} =
	\sqrt{\frac{1}{|\Omega_M|LQ}
	\sum_{(u,v)\in\Omega_M}\sum_{l=1}^{L}\sum_{q=1}^{Q}
	(\Delta_{u,v,l,q})^2},
	$}
	\end{equation}
	where $\Delta$ denotes the prediction error of the corresponding parameter. For angular parameters with periodicity, such as the azimuth angle $\phi$, the circular distance is adopted as
	\vspace{-0.1cm}
	\begin{equation}
		\mbox{\small$\displaystyle
	\Delta\phi=
	\min\left(|\hat{\phi}-\phi|,\;360^\circ-|\hat{\phi}-\phi|\right),
	$}
	\vspace{-0.1cm}
	\end{equation}
	to ensure that the periodic nature of angular predictions is correctly handled.

	\subsubsection{PW-DTW for CIR Evaluation}

	Point-wise error metrics such as RMSE, NMSE, or SSIM are insufficient to capture the \textit{peak misalignment} phenomenon commonly observed in multipath prediction. 
	To jointly evaluate amplitude and delay alignment errors while assigning higher importance to the dominant or strong paths, PW-DTW is introduced to measure the similarity between predicted and ground-truth CIRs.

	\textit{CIR Synthesis and Normalization}:
	For each receiver location, the Top-$L_{\mathrm{pw}}$ dominant propagation paths are selected, with their gains and times of arrival represented as $\{(g_l,\tau_l)\}$. Here, $g_l$ denotes the path gain in dB and $\tau_l$ is measured in ns. The dB gains are first converted to linear amplitudes $p_l = 10^{g_l/10}$. A discrete CIR is then synthesized on a unified delay grid $\{t_i\}_{i=1}^{N_{\mathrm{pw}}}$ using Gaussian pulses as
	\vspace{-0.1cm}
	\begin{equation}
		\mbox{\small$\displaystyle
	x(t_i)=\sum_{l=1}^{L_{\mathrm{pw}}} p_l 
	\exp\!\left(-\frac{(t_i-\tau_l)^2}{2\sigma_g^2}\right),
	$}
	\end{equation}
	where $\sigma_g$ controls the temporal spread of each path. The predicted and ground-truth CIRs are denoted by $y(t_i)$ and $x(t_i)$, respectively, and are jointly normalized using their global maximum amplitude as
	\vspace{-0.1cm}
	\begin{equation}
		\mbox{\small$\displaystyle
	\tilde x(t_i)=\frac{x(t_i)}{\alpha},\quad 
	\tilde y(t_i)=\frac{y(t_i)}{\alpha},
	$}
	\end{equation}
	with
	\begin{equation}
		\mbox{\small$\displaystyle
	\alpha=\max\Big(\max_i x(t_i),\max_i y(t_i)\Big).
	$}
	\vspace{-0.1cm}
	\end{equation}

	\textit{Standard DTW Distance}: 
	Given normalized CIRs, the standard DTW local cost is $c(i,j)=(\tilde x_i-\tilde y_j)^2$, and the cumulative cost matrix is computed via dynamic programming as
	\vspace{-0.1cm}
	\begin{equation}
		\label{G_ij}
	\mbox{\small$\displaystyle
		G(i,j)=c(i,j)+\min\{G(i-1,j),\,G(i,j-1),\,G(i-1,j-1)\},
	$}
	\end{equation}
	yielding the DTW distance $D_{\mathrm{DTW}}=G(N_{\mathrm{pw}},N_{\mathrm{pw}})$, where $G(i,j)$ denotes the minimum cumulative alignment cost between prefixes $\tilde x_{1:i}$ and $\tilde y_{1:j}$. However, treating dominant and weak paths equally fails to prioritize critical propagation alignments.

	\textit{Peak-Weighted Modeling}: 
	To highlight dominant and strong paths, a time-dependent weighting function is constructed based on peak detection. Peak sets $\mathcal{K}_x$ and $\mathcal{K}_y$ are identified on $\tilde x(t)$ and $\tilde y(t)$, respectively, and the Top-$L_{\mathrm{pw}}$ peaks with the largest relative amplitude are retained. Taking $\tilde x(t)$ as an example, the peak-weighting function is defined as
	\vspace{-0.1cm}
	\begin{equation}
		\mbox{\small$\displaystyle
		\begin{aligned}
			W_x(t)&=\max_{k\in \mathcal{K}_x}
			\left(\tilde a_k\exp\!\left(-\frac{(t-t_k)^2}{2\sigma^2}\right)\right),\\
			\tilde a_k&=\frac{\tilde x(t_k)}{\max_i \tilde x(t_i)},
		\end{aligned}
		$}
	\end{equation}
	where $t_k$ denotes the peak location and $\sigma$ controls the influence range of each peak. The function $W_y(t)$ is constructed in the same manner, and both are linearly normalized to $[0,1]$.

	\textit{PW-DTW Weighted Cost and Constraints}: 
	By jointly considering amplitude discrepancies, delay misalignment, and peak importance, the weighted local cost in PW-DTW is expressed as \eqref{cw},
	\begin{figure*}[!htbp]
		\begin{equation}
			\mbox{\small$\displaystyle
			\label{cw}
			\begin{aligned}
				c_w(i,j)=\Big( w_0+\frac{\beta_{\mathrm{pw}}}{2}\big(W_x(t_i)+W_y(t_j)\big)\Big)
				\cdot
				\Big((\tilde x_i-\tilde y_j)^2+\lambda_t(t_i-t_j)^2\Big).
			\end{aligned}
			$}
			\vspace{-0.2cm}
		\end{equation}
		\hrulefill
		\vspace{-0.5cm}
	\end{figure*}
	where $w_0$ is the base weight, $\beta_{\mathrm{pw}}$ controls the strength of peak weighting, and $\lambda_t$ balances the delay penalty.

	To constrain the maximum allowable peak displacement, a Sakoe--Chiba band constraint $|i-j|\le r$ is applied, where $r \approx \Delta t_{\max}/\Delta t$, $\Delta t_{\max}$ denotes the maximum permitted delay offset, and $\Delta t$ is the delay grid resolution. Under this constraint, dynamic programming is performed using $c_w(i,j)$ in place of $c(i,j)$ in \eqref{G_ij} to obtain the PW-DTW distance $D$.

	Since $D$ increases with the length $L_{\mathrm{DTW}}$ of the optimal warping path, a normalized distance is defined as
	\vspace{-0.1cm}
	\begin{equation}
		\mbox{\small$\displaystyle
	\bar D=\frac{D}{L_{\mathrm{DTW}}}.
	$}
	\vspace{-0.1cm}
	\end{equation}
	Averaging $\bar D$ across valid test-set receivers yields the final PW-DTW score, comprehensively evaluating CIR prediction quality regarding peak alignment and amplitude consistency. 

	PW-DTW particularly suits CIR sequences featuring sparse peak structures dominated by a few strong paths. For scenarios involving dense multipath, indistinguishable peaks, or severe noise, increasing the detection threshold or pre-denoising the CIR sequences effectively improves stability.

	\vspace{-0.1cm}
	\section{Experiments and Analysis}
	\label{section5}

	\subsection{Experimental Setup}

	We comprehensively evaluate the proposed framework under cross-scene generation and in-scene completion settings. Given the prohibitive cost and complexity of acquiring dense real-world multipath measurements at scale, relying on RT-based datasets has become the standard practice. High-fidelity RT rigorously simulates electromagnetic propagation to yield accurate multipath parameters, serving as the widely accepted ground truth in related literature \cite{UNet}, \cite{Hoppe2017}.

	\subsubsection{Cross-Scene Generation}
	For the task of RF map generation in unseen environments, the large-scale OpenPathNet dataset \cite{openpathnet} was adopted. Comprising tens of thousands of realistic urban scenes from multiple metropolitan cities, it provides diverse building layouts and environmental structures. Each 128 $\times$ 128 m$^2$ scene includes environmental data and multipath RF ray-tracing measurements sampled at a 1 m resolution. The receiver height is fixed at 1 m, with the transmitter at (0,0,30) m operating at 3.6 GHz.

\begin{figure}[htbp]
		\vspace{-0.4cm}
		\centering
		\subfigure[] {\label{2s1}\centering\includegraphics[width=0.42\columnwidth]{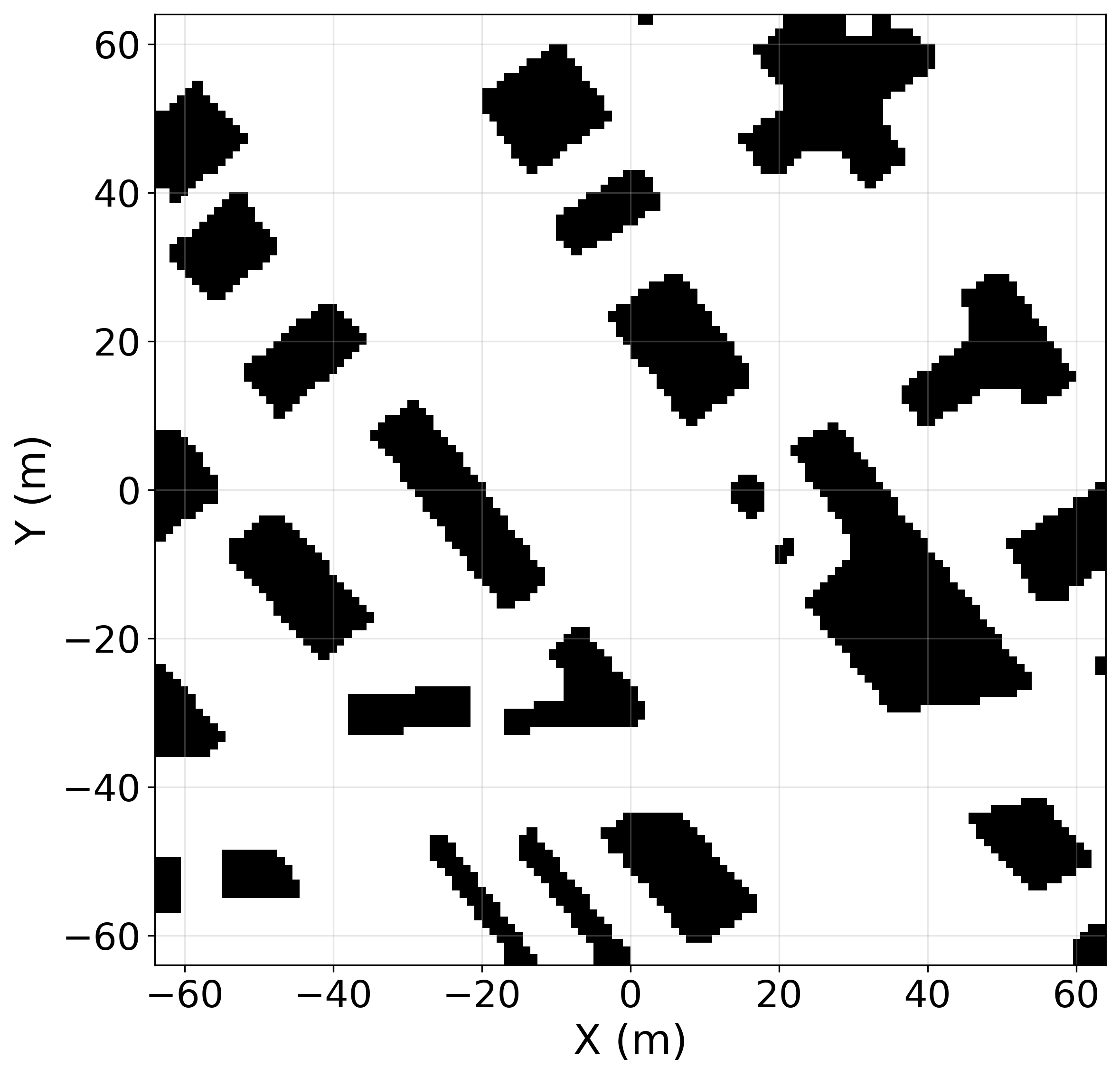}%
		}
		\subfigure[] {\label{2s2}\centering\includegraphics[width=0.49\columnwidth]{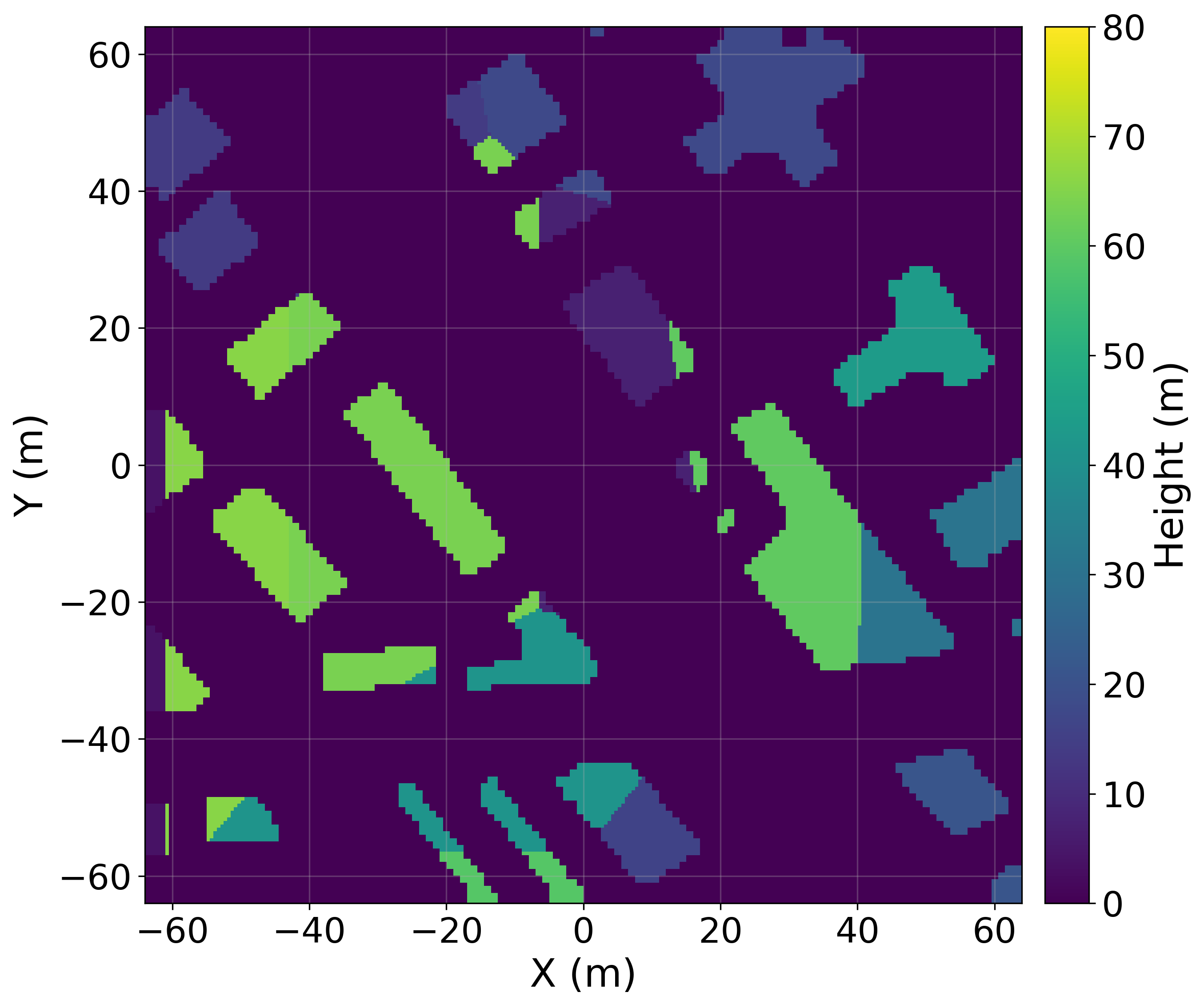}%
		}
		\vspace{-0.2cm}
		\caption{Scene environment map: (a) 2D, (b) 2.5D.}
		\label{fig:2s}
		\vspace{-0.2cm}
	\end{figure}

	Both 2D and 2.5D environmental features were extracted from the environment files provided by the dataset, as illustrated in Fig. \ref{fig:2s}. The model was trained on a subset consisting of $3{,}000$ distinct scenes. To rigorously assess generalization capability, the optimal model checkpoint was selected based on performance on an independent validation set, which was strictly excluded from gradient updates throughout training.

	\subsubsection{In-Scene Completion}
\begin{figure}[htbp]
		\vspace{-0.4cm}
		\centering
		\subfigure[] {\label{DeepMIMO}\centering\includegraphics[width=0.32\columnwidth]{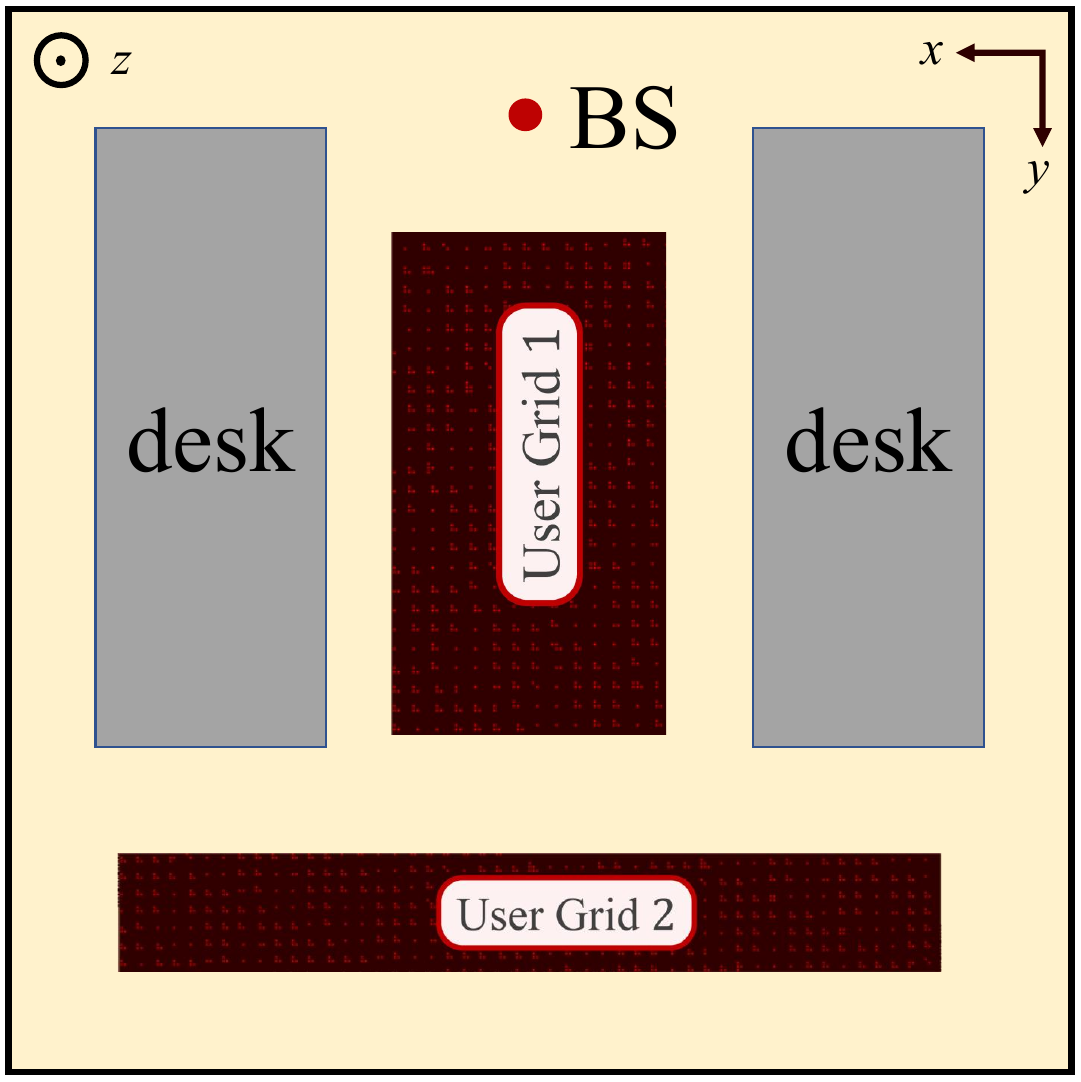}%
		}
		\subfigure[] {\label{USTC1}\centering\includegraphics[width=0.32\columnwidth]{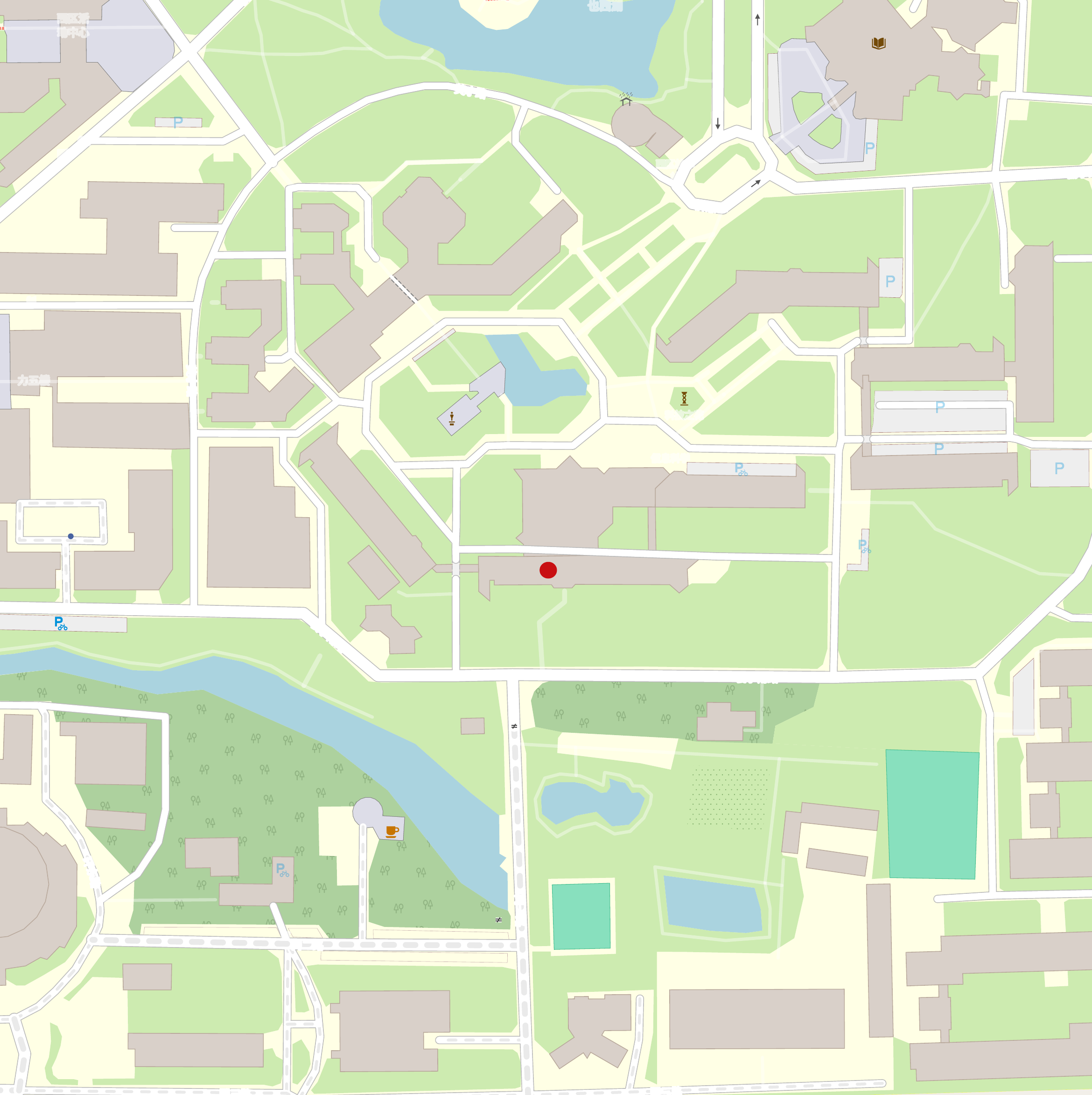}%
		}
		\subfigure[] {\label{USTC2}\centering\includegraphics[width=0.32\columnwidth]{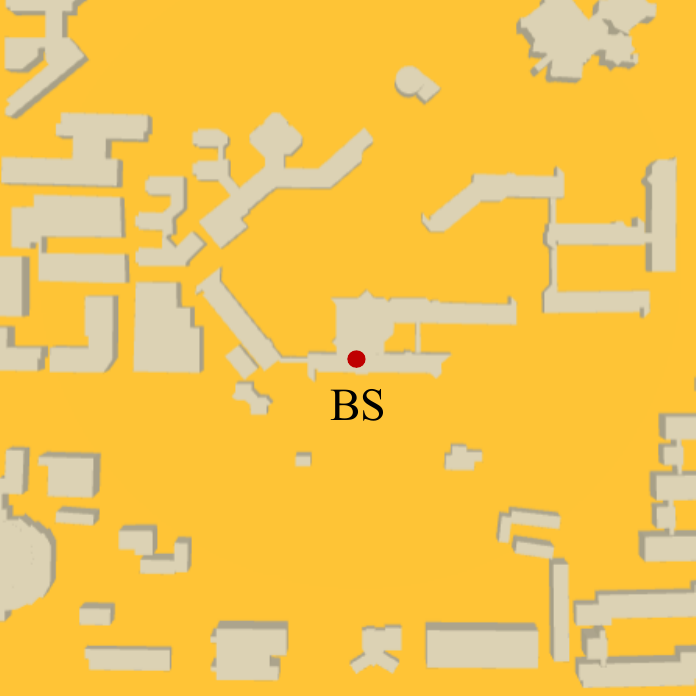}%
		}
		\vspace{-0.2cm}
		\caption{Experimental scenarios: (a) DeepMIMO dataset, (b) USTC campus, (c) 3D model of USTC campus buildings.}
		\label{fig:scenarios}
		\vspace{-0.2cm}
	\end{figure}

	For the task of reconstructing RF maps from sparse observations within a single scene, experiments were conducted on two representative scenarios:

	\begin{itemize}
		\item \textbf{Scenario~1 (Indoor):} This scenario was constructed based on the DeepMIMO dataset \cite{DeepMIMO} and consists of a 10 $\times$ 10 $\times$ 5 m$^3$ indoor environment, as shown in Fig.~\ref{DeepMIMO}. The access point was located at a height of 2.5 m and operated at 2.4 GHz. Within the room, two receiver grids were defined over two designated regions with a spacing of 10 cm, comprising a total of 1,642 receiver locations.
		\item \textbf{Scenario~2 (Outdoor):} This scenario covers a large-scale area of $512 \times 512~\text{m}^2$ on the west campus of the University of Science and Technology of China (USTC), as illustrated in Fig.~\ref{USTC1}. The multipath data were generated using the Sionna ray-tracing platform \cite{Sionna}, following the procedure described in \cite{openpathnet}. Operating at 3.6 GHz, a base station sits 30 m high above the Third Electronics Building, serving receivers uniformly spaced at 1 m intervals at a 1 m height.
	\end{itemize}

	For both completion scenarios, model performance was evaluated under a fixed sparse sampling ratio of 10\%. The randomly selected sparse observation points were divided into training, validation, and test sets with a ratio of $7:1.5:1.5$.

	Since the fifth propagation path's power typically drops to approximately 2\%--5\% of the dominant path, the number of multipath components is set to $L=5$ to balance reconstruction fidelity and computational efficiency. The AdamW optimizer is adopted alongside a validation-driven adaptive decay strategy, using an initial learning rate of $1\times10^{-4}$ globally and $5\times10^{-5}$ specifically for the encoder. To mitigate overfitting, weight decay is set to $1\times10^{-5}$, and a 0.2 dropout rate is applied to both the PINN and GNN modules. Finally, the key hyperparameters are summarized in Table~\ref{tab:hyperparameters} and remain fixed unless otherwise specified in the ablation studies.

	\begin{table}[htbp]
		\centering
		\caption{Key Hyperparameters and Default Settings}
		\label{tab:hyperparameters}
		\setlength{\tabcolsep}{3.5pt}
		\small
		\renewcommand{\arraystretch}{1}
		\begin{tabular}{c c c}
		\toprule
		\textbf{Symbol} & \textbf{Description} & \textbf{Value} \\
		\midrule
		\multicolumn{3}{l}{\textit{Model Architecture}} \\
		$P$ & Number of predicted paths & 5 \\
		$Q$ & Parameters per path & 4 \\
		$k$ & Number of kNN neighbors & 16 \\
		$d_{\text{graph}}$ & GNN hidden dimension & 128 \\
		$L_{\text{graph}}$ & Number of GNN layers & 4 \\
		\midrule
		\multicolumn{3}{l}{\textit{Physics and Regularization}} \\
		$\boldsymbol{\alpha}$ & Power reference coefficients & $[1, 0.7, 0.3, 0.1]$ \\
		$\boldsymbol{\beta}$ & ToA reference coefficients & $[1, 1.2, 1.4, 1.6]$ \\
		$\eta$ & Entropy exponent & 2 \\
		$\epsilon$ & Numerical stability constant & $10^{-8}$ \\
		$\sigma_\angle$ & LoS angular tolerance & $30^\circ$ \\
		$\lambda_{\text{phy}}^{\max}$ & Maximum physics loss weight & 0.5 \\
		$\lambda_{\text{type}}^{\max}$ & Maximum type loss weight & 0.3 \\
		\midrule
		\multicolumn{3}{l}{\textit{PW-DTW Evaluation}} \\
		$K_{\text{pw}}$ & Number of dominant peaks & 5 \\
		$N_{\text{pw}}$ & Delay grid resolution & 5000 \\
		$\sigma_g, \sigma$ & Gaussian pulse width & 3.5 ns \\
		$\beta_{\text{pw}}$ & Peak emphasis factor & 3.0 \\
		$\lambda_t$ & Temporal penalty weight & $1\times 10^{-4}$ \\
		$\Delta t_{\max}$ & Maximum delay offset & 20 ns \\
		\bottomrule
		\end{tabular}
	\end{table}

	\vspace{-0.5cm}
	\subsection{Simulation Results}

	\subsubsection{Performance on Cross-Scene Generation (Map-level)}

	Table~\ref{tab:comparison} reports the quantitative results for the cross-scene generation task. Our unified framework consistently outperforms all image-based baselines across all evaluation metrics. In particular, the proposed 2.5D model achieves the lowest RMSE and NMSE. In terms of structural fidelity, the highest SSIM is obtained by the proposed method, indicating that the physics-guided spatial refinement is more effective at preserving sharp signal discontinuities and shadow boundaries than purely generative models or the regression-based RadioUNet.
	Notably, even the proposed 2D setting yields lower prediction errors than the diffusion-based RadioDiff. This result suggests the performance advantage stems not solely from richer input representations, but largely from the robustness of the integrated PINN-GNN architecture. Finally, the additional performance gains observed in the 2.5D mode confirm that explicitly encoding height information $\mathcal{E}^{2.5\mathrm{D}}$ provides critical geometric cues for modeling diffraction and blockage effects in complex urban environments.

	\begin{table}[htbp]
		\vspace{-0.3cm}
		\centering
		\caption{Map-level Comparison on Cross-Scene Generation}
		\label{tab:comparison}
		\small
		\begin{tabular}{llccc}
			\toprule
			\multicolumn{2}{c}{\textbf{Method}} & \textbf{RMSE} $\downarrow$ & \textbf{NMSE} $\downarrow$ & \textbf{SSIM} $\uparrow$ \\
			\midrule
			\multicolumn{2}{l}{RadioUNet \cite{UNet}} & 0.0237 & 0.0069 & 0.9583 \\
			\multicolumn{2}{l}{RME-GAN \cite{RME_GAN}}     & 0.0296 & 0.0121 & 0.9299 \\
			\multicolumn{2}{l}{RadioDiff \cite{RadioDiff}} & 0.0190 & 0.0049 & 0.9691 \\
			\midrule
			\multirow{2}{*}{\textbf{Proposed}} & 2D        & 0.0188 & 0.0046 & 0.9700 \\
											& \textbf{2.5D}      & \textbf{0.0171} & \textbf{0.0040} & \textbf{0.9725} \\
			\bottomrule
		\end{tabular}
	\end{table}

	\subsubsection{Performance on Cross-Scene Generation (Tensor-level)}

\begin{figure*}[htbp]
		\centering
		\subfigure[] {\label{fig:cross_scene_CDF1}\centering\includegraphics[width=0.4\textwidth]{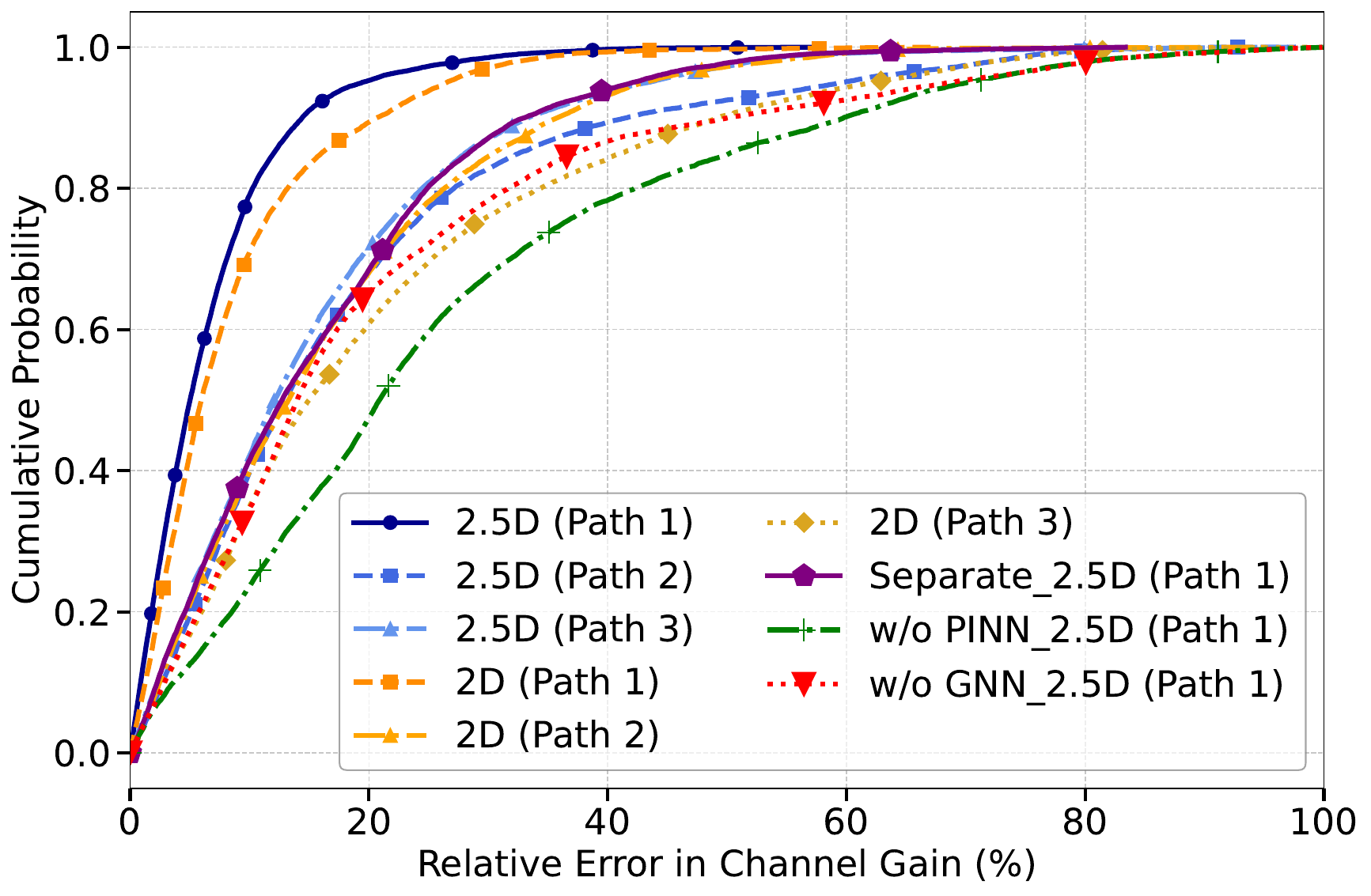}%
		}
		\subfigure[] {\label{fig:cross_scene_CDF2}\centering\includegraphics[width=0.4\textwidth]{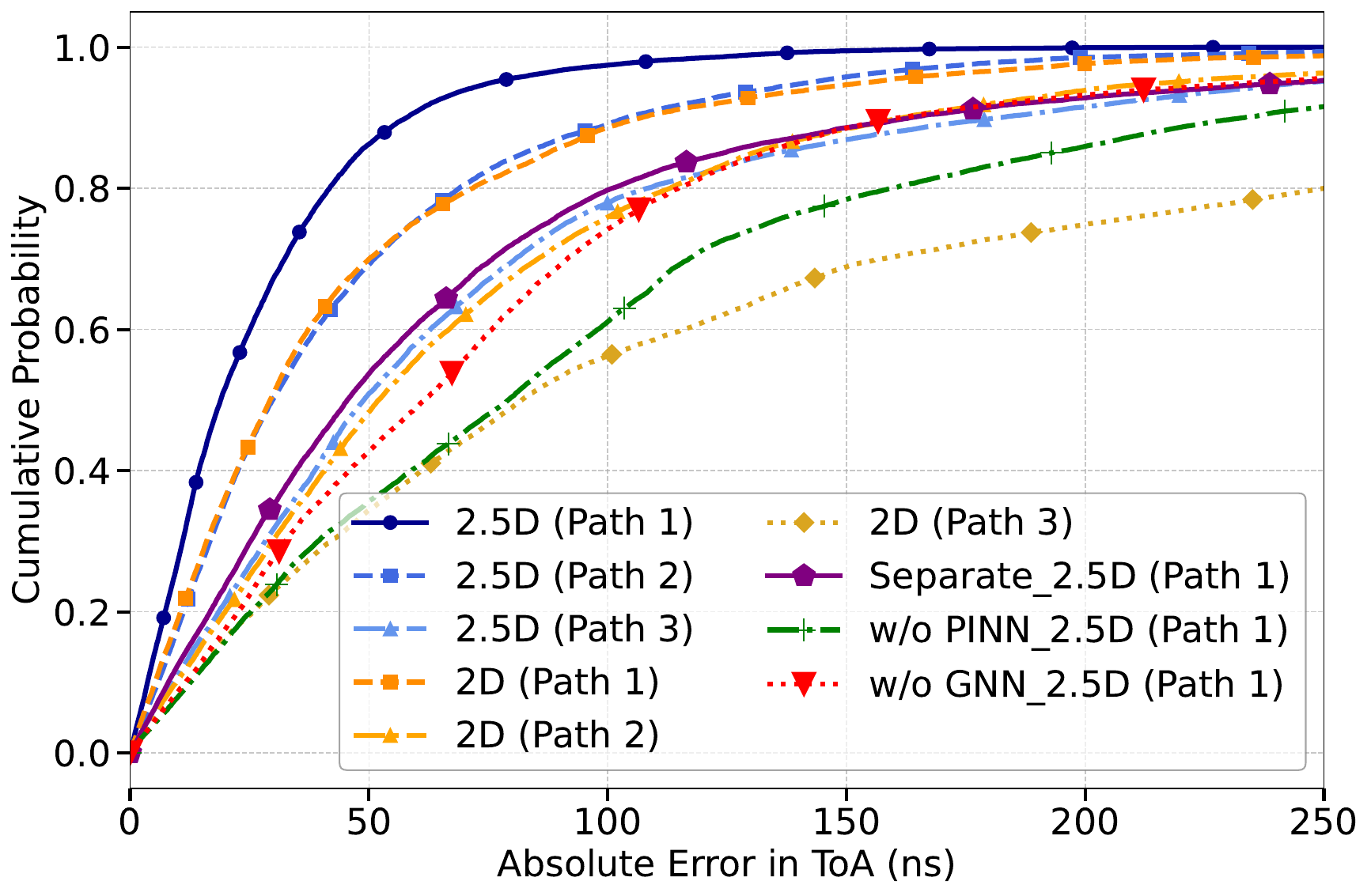}%
		}\\
		\vspace{-0.3cm}
		\subfigure[] {\label{fig:cross_scene_CDF3}\centering\includegraphics[width=0.4\textwidth]{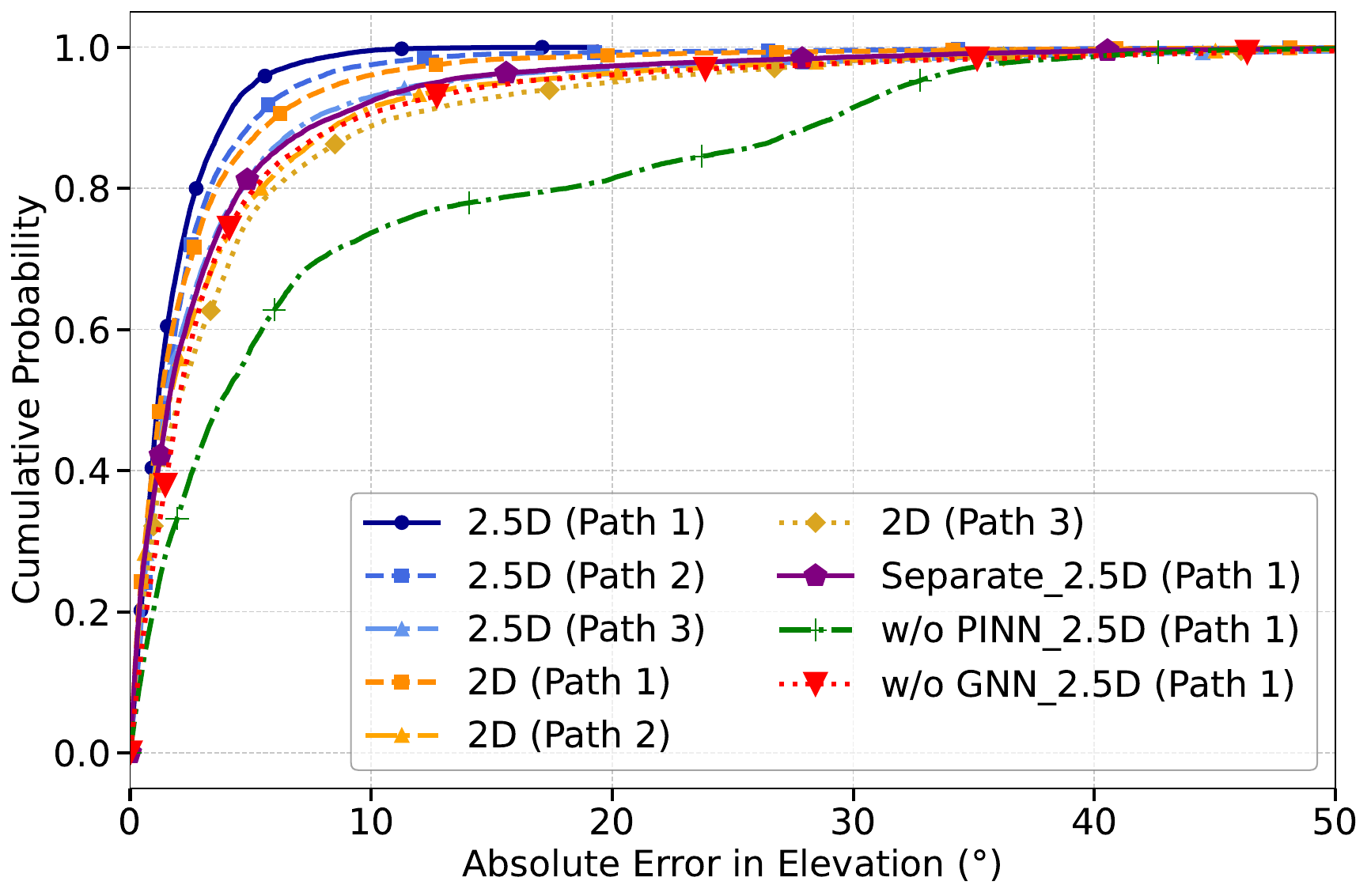}%
		}
		\subfigure[] {\label{fig:cross_scene_CDF4}\centering\includegraphics[width=0.4\textwidth]{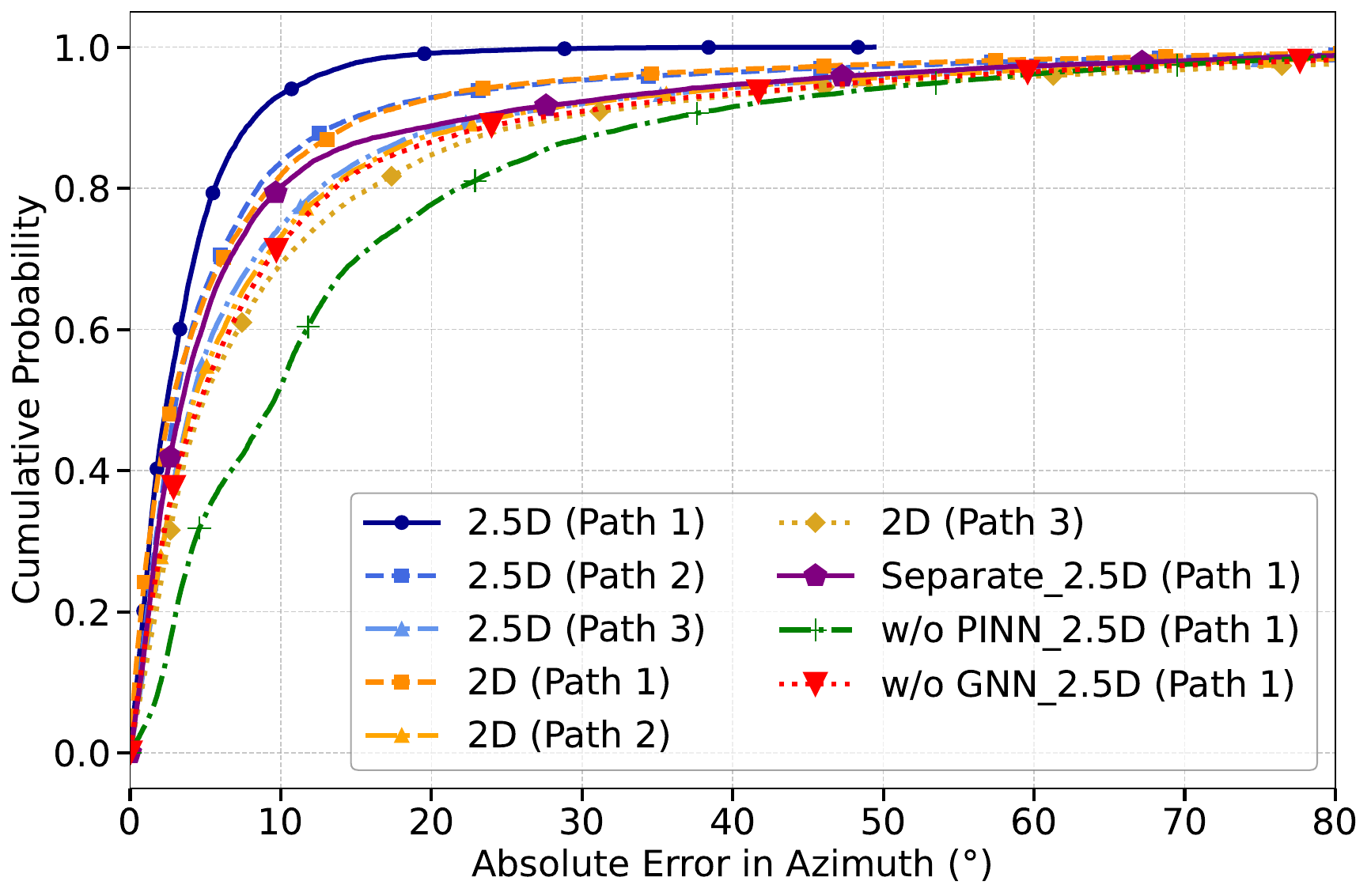}%
		}
		\vspace{-0.2cm}
		\caption{CDFs of multipath prediction errors for the cross-scene generation task: (a) channel gain, (b) ToA, (c) elevation angle, and (d) azimuth angle.}
		\label{fig:cross_scene_CDF}
		\vspace{-0.5cm}
	\end{figure*}

	Fig.~\ref{fig:cross_scene_CDF} presents the cumulative distribution function (CDF) curves of the prediction errors for four key physical parameters across 2D and 2.5D input configurations alongside ablation variants. Owing to the large dynamic range among path gains, path-gain error is evaluated as a relative error. The proposed 2.5D method consistently achieves the best first-path prediction performance across all parameters, significantly outperforming the 2D configuration and all ablated variants. Under the 2.5D setting, approximately 95\% of first-path gain predictions exhibit relative errors below 20\%, while 80\% of ToA errors fall within 50 ns. Elevation and azimuth angle errors are largely constrained within $10^\circ$ and $20^\circ$, respectively, demonstrating that explicit height information effectively improves modeling accuracy for complex geometric relationships.
	In contrast, the w/o PINN variant exhibits pronounced long-tail behavior in ToA and angular predictions, substantially increasing the proportion of large-error samples. This validates the critical role of physics-based constraints in preventing parameter divergence. Moreover, as path order increases, enhanced multipath randomness complicates the prediction task, slowing CDF convergence. Even in this regime, the proposed 2.5D method maintains competitive performance, highlighting its robustness in capturing secondary and weak multipath components.

	\begin{table}[htbp]
		\vspace{-0.3cm}
		\centering
		\renewcommand{\arraystretch}{1.1}
		\setlength{\tabcolsep}{3.2pt}
		\caption{Cross-Scene Ablation Results under 2D and 2.5D Environment Representations}
		\label{tab:cross_scene_ablation}
		\small
		\begin{tabular}{c|cc|cc|cc}
		\toprule
		\multirow{2}{*}{Method} 
		& \multicolumn{2}{c|}{RMSE $\downarrow$} 
		& \multicolumn{2}{c|}{NMSE $\downarrow$}
		& \multicolumn{2}{c}{PW-DTW ${\scriptscriptstyle (\times 10^{-3})}$ $\downarrow$} \\
		\cmidrule(lr){2-3} \cmidrule(lr){4-5} \cmidrule(lr){6-7}
		& 2D & 2.5D & 2D & 2.5D & 2D & 2.5D \\
		\midrule
		w/o PINN 
		& 2.0923 & 1.9879 
		& 0.0657 & 0.0597 
		& 3.102 & 2.351 \\

		w/o GNN 
		& 1.8986 & 1.7029 
		& 0.0519 & 0.0486 
		& 2.317 & 1.958 \\

		Separate 
		& 1.7982 & 1.5657 
		& 0.0497 & 0.0454 
		& 2.330 & 1.946 \\

		Proposed 
		& \textbf{1.3877} & \textbf{1.2714} 
		& \textbf{0.0352} & \textbf{0.0324} 
		& \textbf{1.642} & \textbf{1.353} \\
		\bottomrule
		\end{tabular}
		\vspace{-0.3cm}
	\end{table}

	\begin{figure}[htbp]
		\centering
		\subfigure[] {\label{fig:cross_scene_cir1}\centering\includegraphics[width=0.94\columnwidth]{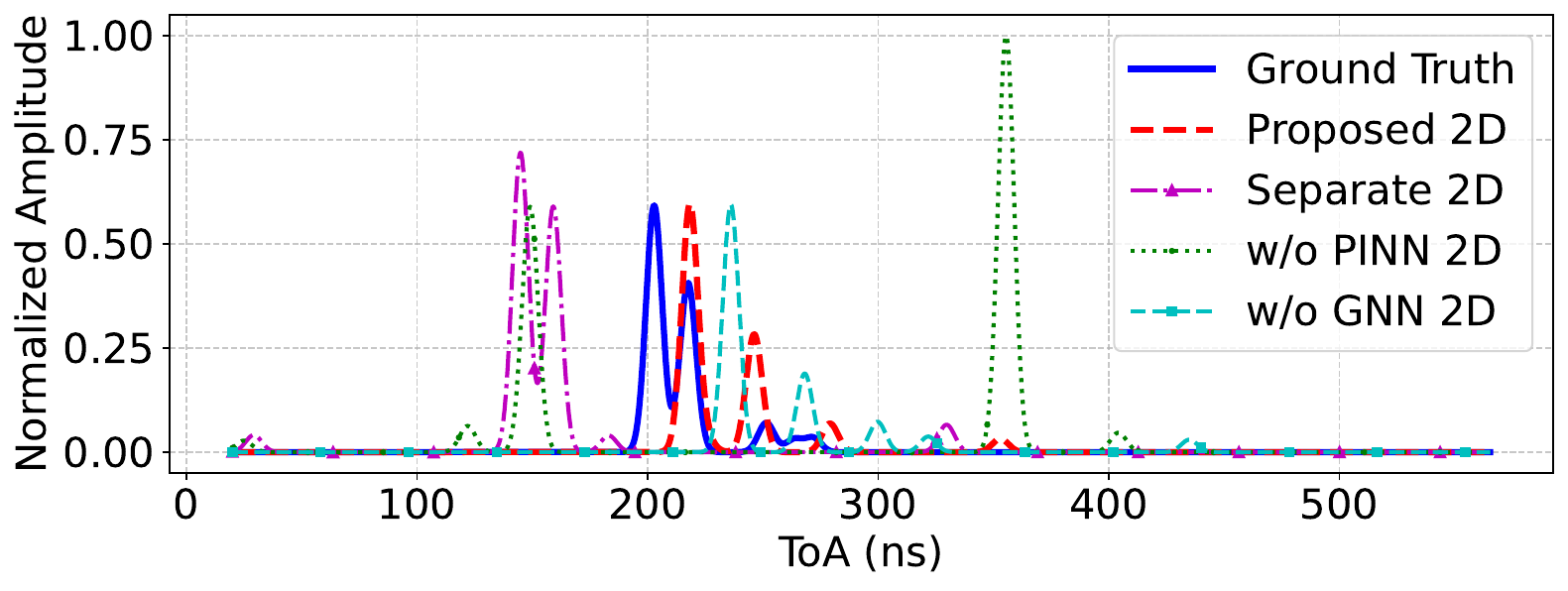}%
		}\\
		\vspace{-0.3cm}
		\subfigure[] {\label{fig:cross_scene_cir2}\centering\includegraphics[width=0.94\columnwidth]{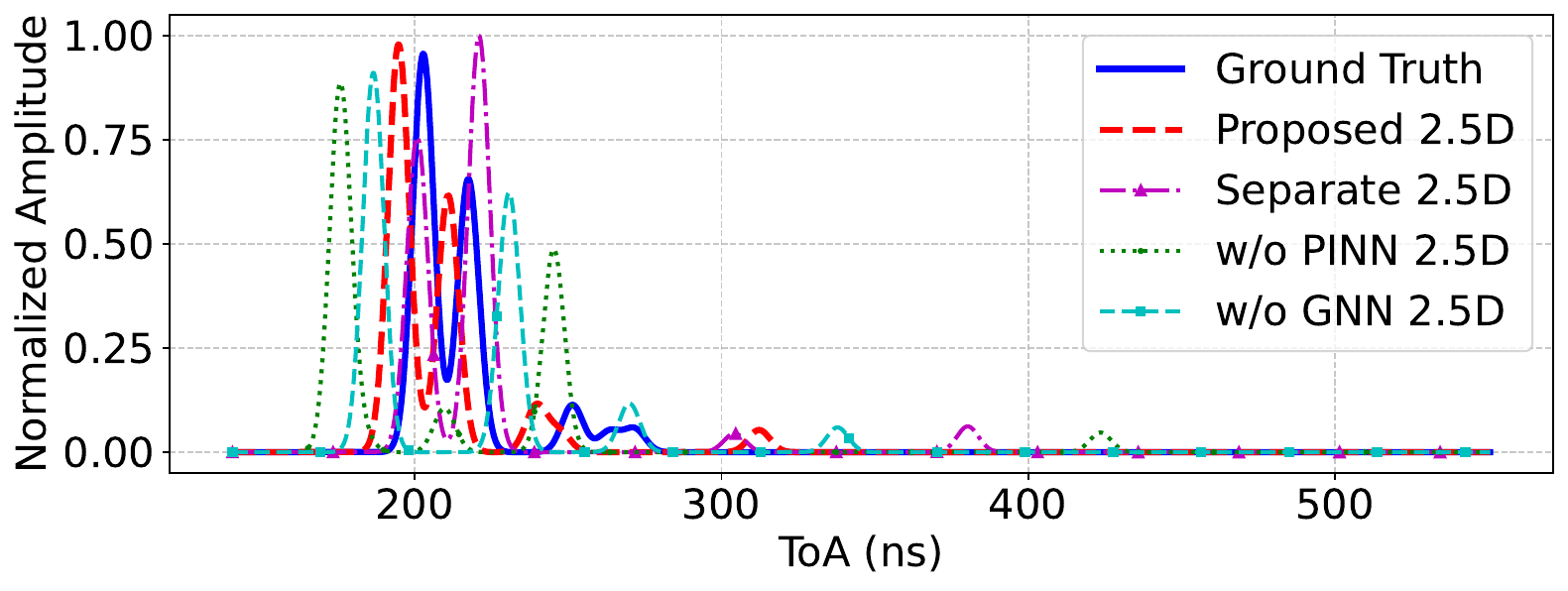}%
		}
		\vspace{-0.3cm}
		\caption{CIR of cross-scene generation scheme: (a) 2D, and (b) 2.5D.}
		\label{fig:cross_scene_cir}
		\vspace{-0.7cm}
	\end{figure}
	
	Table~\ref{tab:cross_scene_ablation} details the cross-scene generation ablation results, quantitatively evaluating how architectural components and input dimensionalities impact multipath parameter prediction. The experimental results indicate that the proposed 2.5D framework achieves the best performance across all evaluation metrics, thereby validating the effectiveness of jointly incorporating physics-based constraints and graph-based reasoning. Specifically, removing physical constraints (w/o PINN) causes the severest degradation, yielding the highest RMSE and NMSE. This underscores that purely data-driven regression struggles to model complex physical parameters across wide dynamic ranges accurately. Furthermore, the Separate scheme highlights the necessity of end-to-end joint optimization, as decoupled PINN and GNN training fails to fully exploit their complementary benefits.

	The PW-DTW metric and qualitative results in Fig.~\ref{fig:cross_scene_cir} further corroborate these conclusions. Under the 2D setting, the w/o PINN variant's sharp PW-DTW error increase directly corresponds to the severe peak misalignment shown by the green dotted curve in Fig.~\ref{fig:cross_scene_cir1}. This confirms the physics loss's decisive role in anchoring ToA predictions. Conversely, the proposed method minimizes numerical errors while generating power delay spectra whose peak locations and amplitudes closely track the ground truth. Furthermore, the 2.5D model consistently outperforms its 2D counterpart across all metrics, demonstrating that explicit height information provides critical diffraction-related geometric cues to enhance overall multipath prediction accuracy.

	\subsubsection{In-Scene Completion Results}


	\begin{figure*}
		\centering
		\subfigure[] {\label{fig:CDF1}\centering\includegraphics[width=0.379\textwidth]{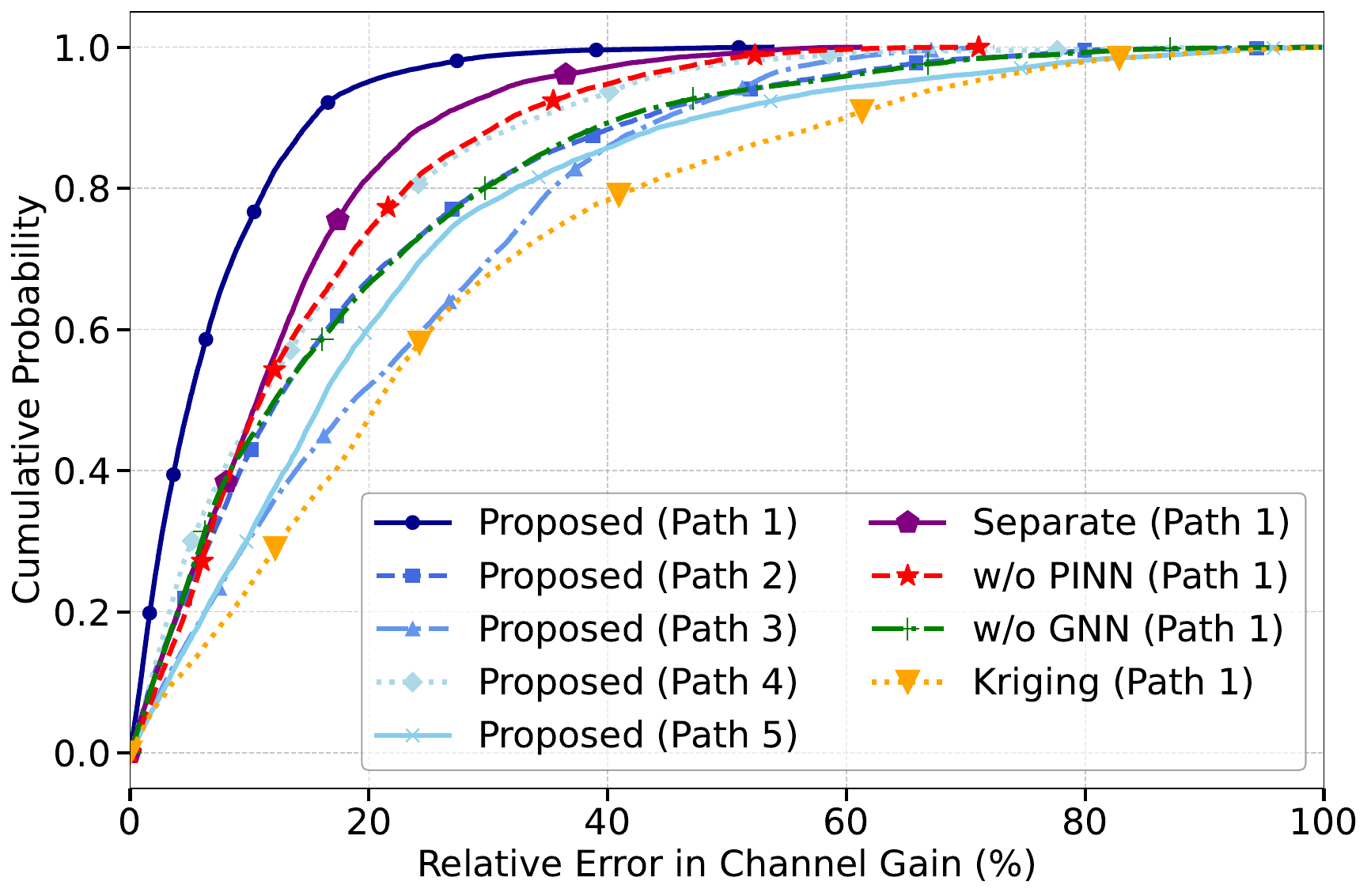}%
		}
		\subfigure[] {\label{fig:CDF2}\centering\includegraphics[width=0.379\textwidth]{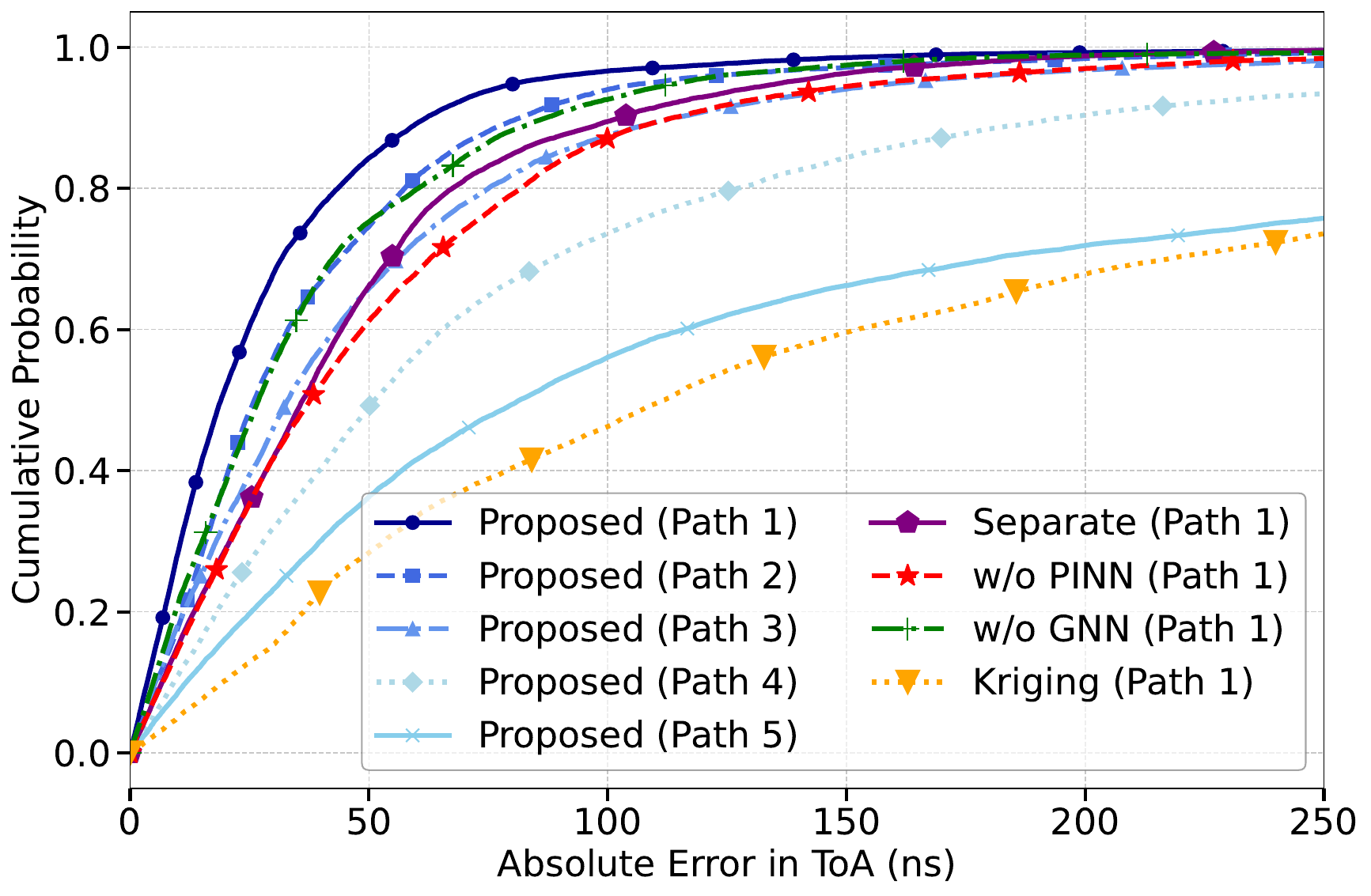}%
		}\\
		\vspace{-0.3cm}
		\subfigure[] {\label{fig:CDF3}\centering\includegraphics[width=0.379\textwidth]{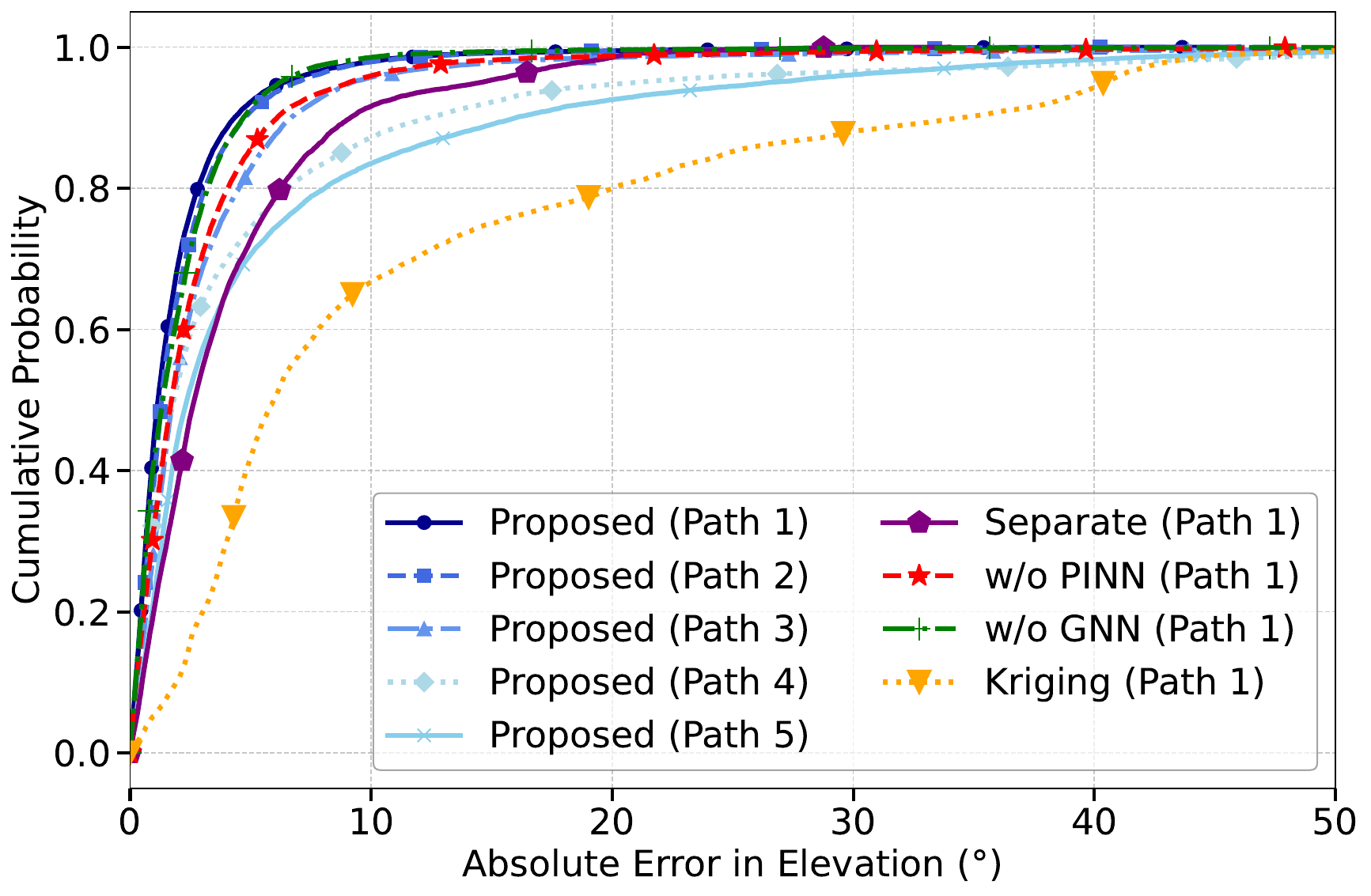}%
		}
		\subfigure[] {\label{fig:CDF4}\centering\includegraphics[width=0.379\textwidth]{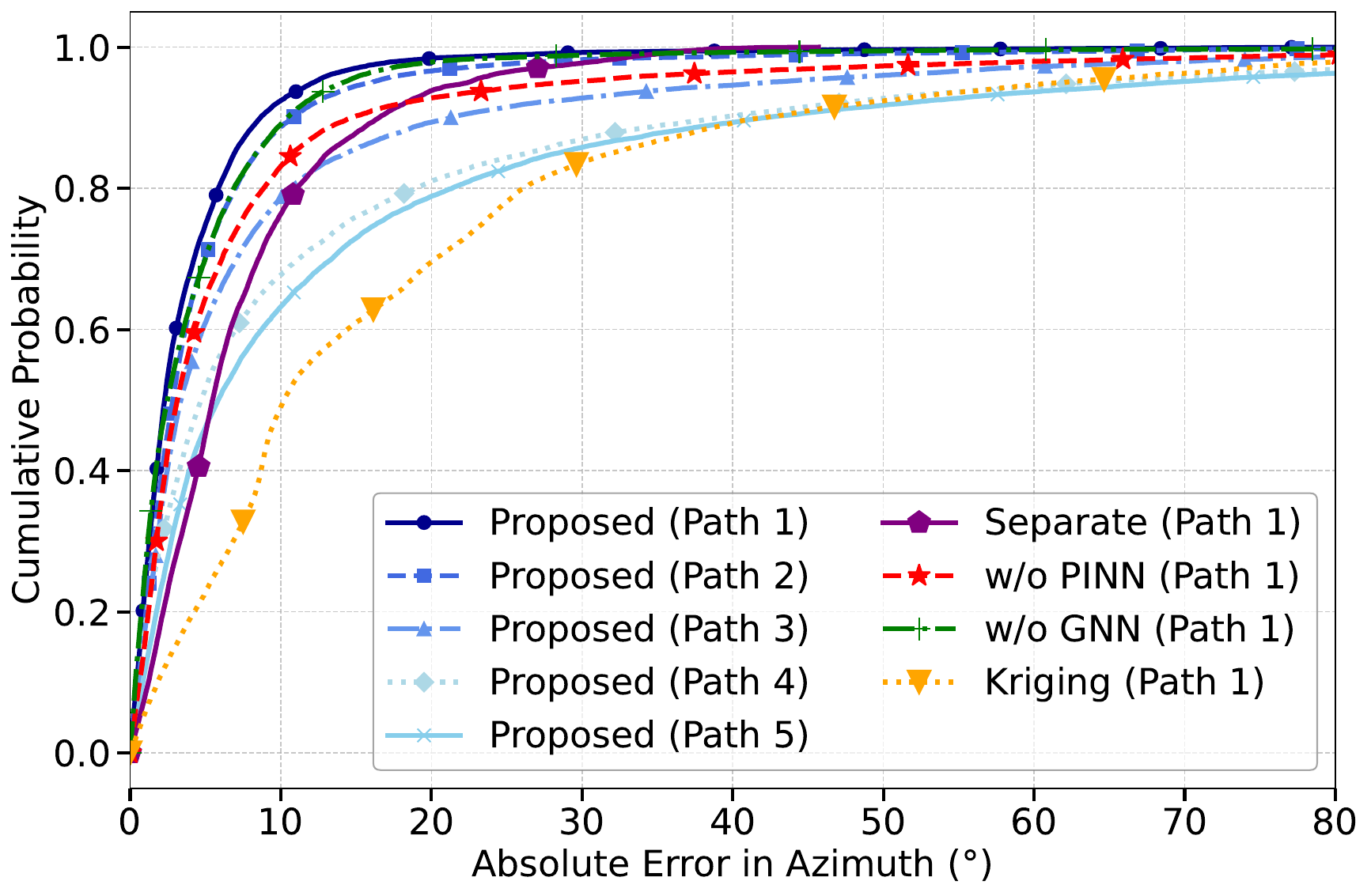}%
		}
		\vspace{-0.3cm}
		\caption{CDFs of multipath prediction errors for the in-scene completion task in S2: (a) channel gain, (b) ToA, (c) elevation angle, and (d) azimuth angle.}
		\label{fig:CDF}
		\vspace{-0.5cm}
	\end{figure*}

	Fig.~\ref{fig:CDF} quantifies these performance gains through cumulative distribution function curves. The proposed method significantly outperforms all baselines in first-path prediction, whereas Kriging exhibits pronounced ToA and angle estimation deficiencies. Specifically, approximately 90\% of first-path gain relative errors fall below 20\%, 80\% of ToA errors remain within 50 ns, and most azimuth errors are confined within $10^\circ$. Furthermore, while enhanced multipath randomness complicates higher-order path predictions, the PINN-GNN framework successfully bounds these errors, validating its robustness in resolving multipath components across complex environments.

	\begin{table}[htbp]
		\vspace{-0.3cm}
		\centering
		\setlength{\tabcolsep}{3.2pt} 
		\renewcommand{\arraystretch}{1}
		\caption{Performance Comparison on In-Scene Completion}
		\label{tab:comparison_vertical}
		\small
		\begin{tabular}{c|cc|cc|cc}
		\toprule
		\multirow{2}{*}{Method} 
		& \multicolumn{2}{c|}{RMSE $\downarrow$} 
		& \multicolumn{2}{c|}{NMSE $\downarrow$}
		& \multicolumn{2}{c}{PW-DTW ${\scriptscriptstyle (\times 10^{-3})}$ $\downarrow$} \\
		\cmidrule(lr){2-3} \cmidrule(lr){4-5} \cmidrule(lr){6-7}
		& S1 & S2 & S1 & S2 & S1 & S2 \\
		\midrule
		Kriging 
		& 2.3166 & 2.6835 
		& 0.0786 & 0.0914
		& 2.673 & 3.445 \\

		w/o PINN 
		& 1.8952 & 2.2473 
		& 0.0566 & 0.0673 
		& 2.260 & 2.398 \\

		w/o GNN 
		& 1.6133 & 1.9072 
		& 0.0460 & 0.0542
		& 1.803 & 2.356 \\

		Separate 
		& 1.5027 & 1.7492 
		& 0.0437 & 0.0501
		& 1.818 & 2.232 \\

		Proposed 
		& \textbf{1.2106} & \textbf{1.3522} 
		& \textbf{0.0317} & \textbf{0.0363} 
		& \textbf{1.287} & \textbf{1.862} \\
		\bottomrule
		\end{tabular}
		\vspace{-0.3cm}
	\end{table}

	\begin{figure}[htbp]
		\centering
		\subfigure[] {\label{fig:cir1}\centering\includegraphics[width=0.94\columnwidth]{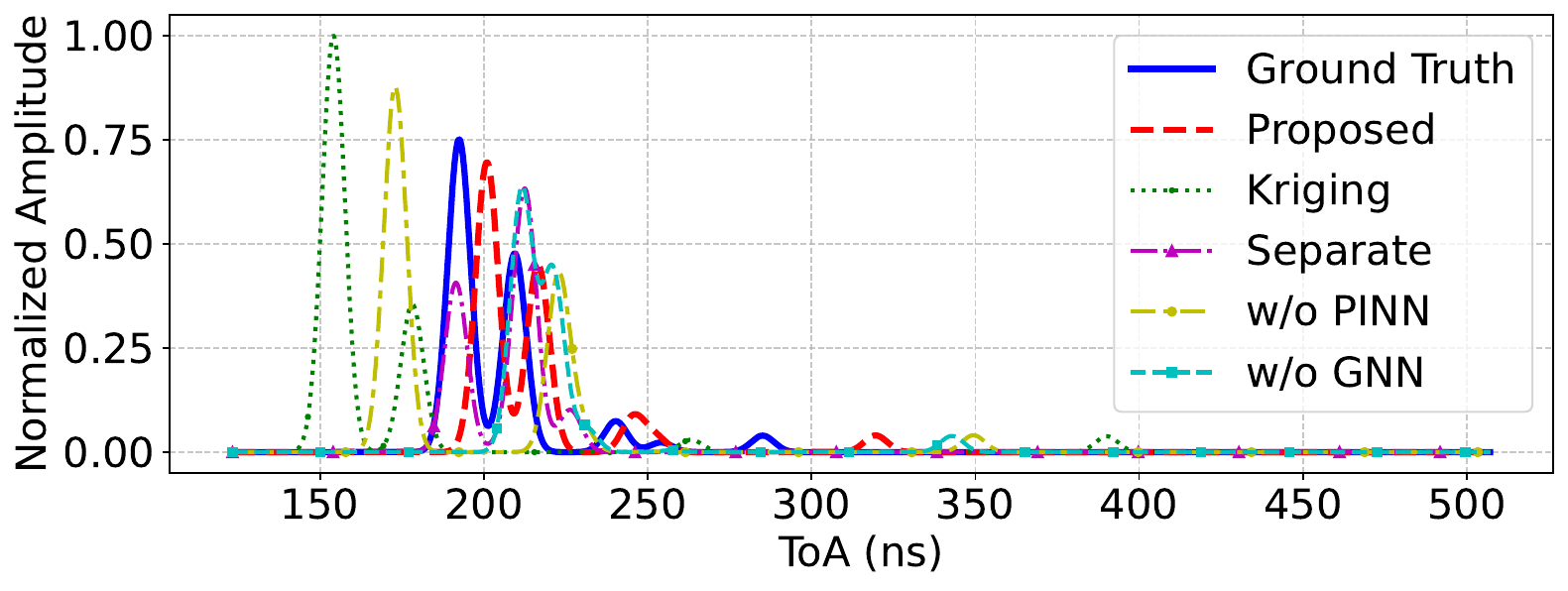}%
		}\\
		\vspace{-0.3cm}
		\subfigure[] {\label{fig:cir2}\centering\includegraphics[width=0.94\columnwidth]{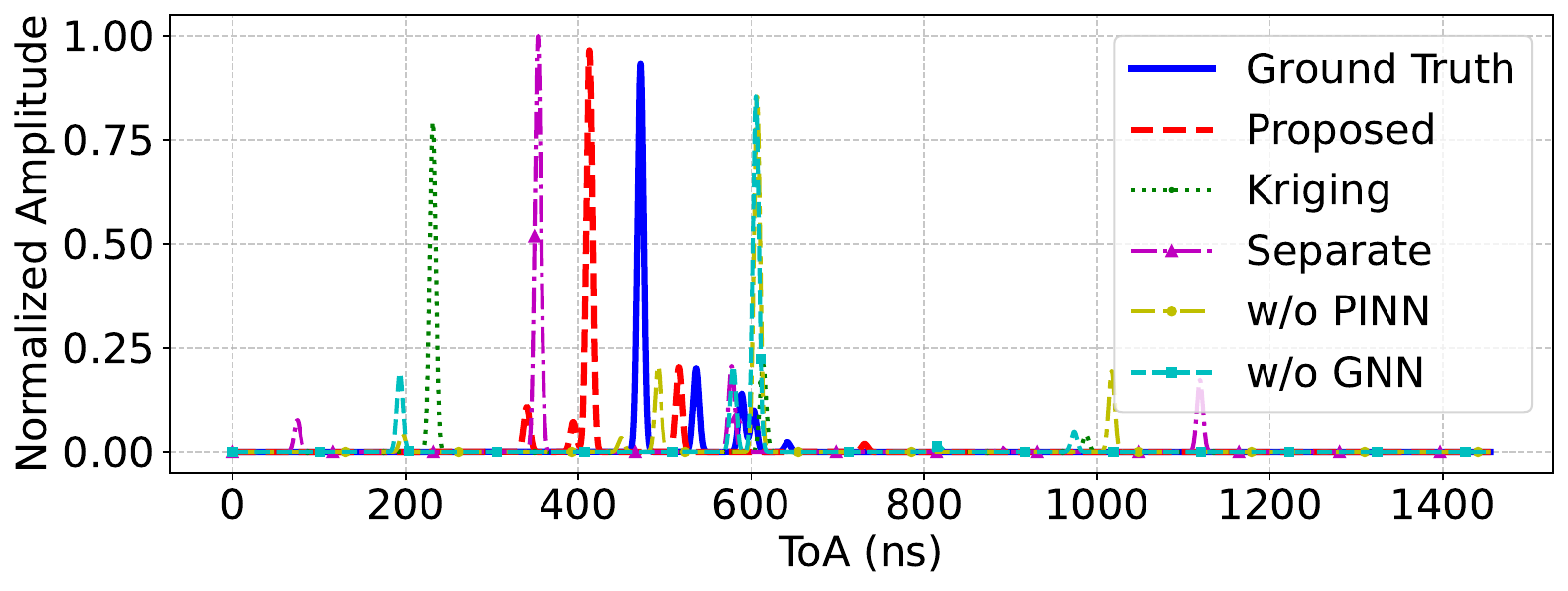}%
		}
		\vspace{-0.3cm}
		\caption{CIR of in-scene completion scheme: (a) S1, and (b) S2.}
		\label{fig:cir}
		\vspace{-0.7cm}
	\end{figure}

	Table~\ref{tab:comparison_vertical} details the in-scene completion performance across Scenarios~1 and~2. The PINN-GNN framework consistently outperforms the Kriging baseline and all ablation variants across all metrics. Particularly in the challenging Scenario~2, the proposed method reduces RMSE by approximately 50\% against Kriging, highlighting the superiority of physics-aware propagation learning over purely statistical interpolation.
	Further insights can be obtained by jointly examining the PW-DTW metric and the power delay profiles shown in Fig.~\ref{fig:cir}. 
	Yielding the highest PW-DTW errors in both scenarios, Kriging produces severely misaligned spurious peaks, as visualized by the green dotted curve in Fig.~\ref{fig:cir1}. This failure stems from a lack of physical awareness regarding light-speed wave propagation and environmental geometry. Similarly, removing physical constraints (w/o PINN) severely degrades PW-DTW performance, confirming the physics loss's critical role in anchoring ToA predictions within physically plausible distances. Conversely, the proposed method minimizes PW-DTW scores, generating curves that closely track the ground truth to capture dominant path delays and multipath structures accurately. Ultimately, these results validate the jointly optimized PINN-GNN framework's effectiveness in reconstructing high-fidelity channel fingerprints.

	Note that the reported RMSE/NMSE values are not directly comparable to those obtained in our earlier work \cite{PINN+GNN}, as the current evaluation adopts per-path and per-feature normalization together with tensor-wise averaging.

	\vspace{-0.3cm}
	\section{Conclusion}
	\label{section6}

	In this paper, we propose a unified PINN-GNN framework for constructing high-fidelity multipath radio maps in complex urban environments, supporting both cross-scenario generation and sparse-sampling completion using 2D and 2.5D representations. By embedding physical constraints, the PINN learns a physically consistent mapping from receiver locations to multipath parameters, while the GNN captures spatial correlations via a $k$-nearest-neighbor graph to refine predictions. The PINN-GNN framework jointly models the physical relationships among multipath components and their spatial consistency, enabling accurate characterization of multipath propagation.
	Extensive experimental results demonstrate that the proposed method exhibits strong generalization performance in both cross-scenario generation and in-scenario completion tasks, achieving consistently superior map-level metrics compared to relevant baselines. Notably, under sparse sampling, the PINN-GNN framework maintains significantly higher robustness than Kriging interpolation. Through ablation studies and the PW-DTW metric, we further verify the decisive role of physical constraints in anchoring time-domain signal characteristics. These results validate the effectiveness of integrating physical knowledge with deep learning, and provide a reliable solution for high-precision digital twins and real-time channel prediction in future 6G networks.

\vspace{-0.3cm}
\bibliographystyle{IEEEtran}
\bibliography{Refs}

@inproceedings{PINN+GNN,
  address = {Belgrade, Serbia},
  author={Liu, Lizhou and Chen, Xiaohui and Tang, Zihan and Ma, Mengyao and Zhang, Wenyi},
  booktitle={2025 International Conference on Future Communications and Networks (FCN)}, 
  title={{PINN} and {GNN}-based {RF} Map Construction for Wireless Communication Systems}, 
  month={Aug.},
  year={2025},
  volume={},
  number={},
  pages={1-6},
}

@article{6G_1,
  author  = {Rafique, W. and Barai, J. and Fapojuwo, A. O. and Krishnamurthy, D.},
  title   = {A survey on beyond {5G} network slicing for smart cities applications},
  journal = {IEEE Commun. Surveys Tuts.},
  volume  = {27},
  number  = {1},
  pages   = {595--628},
  month   = {Feb.},
  year    = {2025}
}

@article{Sun2025,
  author  = {Sun, R. and others},
  title   = {A comprehensive survey of knowledge-driven deep learning for intelligent wireless network optimization in {6G}},
  journal = {IEEE Commun. Surveys Tuts.},
  volume  = {28},
  pages   = {1099--1135},
  month   = {May},
  year    = {2025}
}

@article{Wang2024,
  author  = {Wang, Z. and others},
  title   = {A tutorial on extremely large-scale {MIMO} for {6G}: Fundamentals, signal processing, and applications},
  journal = {IEEE Commun. Surveys Tuts.},
  volume  = {26},
  number  = {3},
  pages   = {1560--1605},
  month   = {3rd Quart.,},
  year    = {2024}
}

@article{Ma2024,
  author  = {Ma, T. and Qian, B. and Qin, X. and Liu, X. and Zhou, H. and Zhao, L.},
  title   = {Satellite-terrestrial integrated {6G}: An ultra-dense {LEO} networking management architecture},
  journal = {IEEE Wireless Commun.},
  volume  = {31},
  number  = {1},
  pages   = {62--69},
  month   = {Feb.},
  year    = {2024}
}

@article{Zeng2024,
  author  = {Zeng, Y. and others},
  title   = {A tutorial on environment-aware communications via channel knowledge map for {6G}},
  journal = {IEEE Commun. Surveys Tuts.},
  volume  = {26},
  number  = {3},
  pages   = {1478--1519},
  month   = {3rd Quart.,},
  year    = {2024}
}

@article{Okumura,
  author  = {Okumura, Y.},
  title   = {Field strength and its variability in {VHF} and {UHF} land-mobile radio service},
  journal = {Rev. Electr. Commun. Lab.},
  volume  = {16},
  pages   = {825--873},
  year    = {1968}
}

@article{Hata,
  author  = {Hata, M.},
  title   = {Empirical formula for propagation loss in land mobile radio services},
  journal = {IEEE Trans. Veh. Technol.},
  volume  = {29},
  number  = {3},
  pages   = {317--325},
  month   = {Aug.},
  year    = {1980}
}

@techreport{3GPP,
  author      = {{3GPP}},
  title       = {{3GPP TR38.901}},
  institution = {3rd Generation Partnership Project (3GPP)},
  url         = {https://portal.3gpp.org/desktopmodules/Specifications/SpecificationDetails.aspx?specificationId=3173},
  year        = {2023}
}

@article{Rizk1997,
  author  = {Rizk, K. and Wagen, J.-F. and Gardiol, F.},
  title   = {Two-dimensional ray-tracing modeling for propagation prediction in microcellular environments},
  journal = {IEEE Trans. Veh. Technol.},
  volume  = {46},
  number  = {2},
  pages   = {508--518},
  month   = {May},
  year    = {1997}
}

@article{Sionna,
  author  = {Hoydis, J. and Aoudia, F. A. and Cammerer, S. and Nimier-David, M. and Binder, N. and Marcus, G. and Keller, A.},
  title   = {{Sionna RT}: Differentiable ray tracing for radio propagation modeling},
  journal = {arXiv preprint arXiv:2303.11103},
  year    = {2023}
}

@article{Kriging,
  author  = {Dall'Anese, E. and Kim, S.-J. and Giannakis, G. B.},
  title   = {Channel gain map tracking via distributed {Kriging}},
  journal = {IEEE Trans. Veh. Technol.},
  volume  = {60},
  number  = {3},
  pages   = {1205--1211},
  month   = {Mar.},
  year    = {2011}
}

@inproceedings{Chouvardas2016,
  author    = {Chouvardas, S. and Valentin, S. and Draief, M. and Leconte, M.},
  title     = {A method to reconstruct coverage loss maps based on matrix completion and adaptive sampling},
  booktitle = {Proc. {IEEE} Int. Conf. Acoust., Speech Signal Process. ({ICASSP})},
  address   = {Shanghai, China},
  pages     = {6390--6394},
  month     = {Mar.},
  year      = {2016}
}

@inproceedings{Schaufele2019,
  author    = {Schaufele, D. and Cavalcante, R. L. G. and Stanczak, S.},
  title     = {Tensor completion for radio map reconstruction using low rank and smoothness},
  booktitle = {Proc. {IEEE} 20th Int. Workshop Signal Process. Adv. Wireless Commun. ({SPAWC})},
  address   = {Cannes, France},
  pages     = {1--5},
  month     = {Jul.},
  year      = {2019}
}

@article{UNet,
  author  = {Levie, R. and Yapar, C. and Kutyniok, G. and Caire, G.},
  title   = {{RadioUNet}: fast radio map estimation with convolutional neural networks},
  journal = {IEEE Trans. Wireless Commun.},
  volume  = {20},
  pages   = {4001--4015},
  month   = {Jun.},
  year    = {2021}
}

@inproceedings{Hoppe2017,
  author    = {Hoppe, R. and Wolfle, G. and Jakobus, U.},
  title     = {Wave propagation and radio network planning software {WinProp} added to the electromagnetic solver package {FEKO}},
  booktitle = {Proc. Int. Appl. Comput. Electromagn. Soc. Symp. Italy (ACES)},
  address   = {Florence, Italy},
  pages     = {1--2},
  month     = {Mar.},
  year      = {2017}
}

@inproceedings{WNet,
  author    = {Li, Y. and Li, Z. and Gao, Z. and Chen, T.},
  title     = {{Geo2SigMap}: High-fidelity {RF} signal mapping using geographic databases},
  booktitle = {2024 IEEE International Symposium on Dynamic Spectrum Access Networks (DySPAN)},
  pages     = {277--285},
  month     = {May},
  year      = {2024}
}

@article{Chen2023,
  author  = {Chen, G. and Liu, Y. and Zhang, T. and Zhang, J. and Guo, X. and Yang, J.},
  title   = {A graph neural network based radio map construction method for urban environment},
  journal = {IEEE Commun. Lett.},
  volume  = {27},
  number  = {5},
  pages   = {1327--1331},
  month   = {May},
  year    = {2023}
}

@article{RME_GAN,
  author  = {Zhang, S. and Wijesinghe, A. and Ding, Z.},
  title   = {{RME-GAN}: A learning framework for radio map estimation based on conditional generative adversarial network},
  journal = {IEEE Internet Things J.},
  volume  = {10},
  number  = {20},
  pages   = {18016-18027},
  month   = {Oct.},
  year    = {2023}
}

@article{Chen2024,
  author  = {Chen, Q. and Yang, J. and Huang, M. and Zhou, Q.},
  title   = {{ACT-GAN}: Radio map construction based on generative adversarial networks with {ACT} blocks},
  journal = {IET Commun.},
  volume  = {18},
  number  = {19},
  pages   = {1541--1550},
  month   = {Dec.},
  year    = {2024}
}

@inproceedings{Sarkar2024,
  author    = {Sarkar, S. and Manshaei, M. H. and Krunz, M. and Ravaee, H.},
  title     = {{RecuGAN}: A novel generative {AI} approach for synthesizing {RF} coverage maps},
  booktitle = {33rd International Conference on Computer Communications and Networks (ICCCN)},
  address   = {Kailua-Kona, HI, USA},
  pages     = {1--9},
  year      = {2024}
}

@article{RadioDiff,
  author  = {Wang, X. and Tao, K. and Cheng, N. and Yin, Z. and Li, Z. and Zhang, Y. and Shen, X.},
  title   = {{RadioDiff}: An effective generative diffusion model for sampling-free dynamic radio map construction},
  journal = {IEEE Trans. Cogn. Commun. Netw.},
  volume  = {11},
  number  = {2},
  pages   = {738--750},
  month   = {Apr.},
  year    = {2025}
}

@article{Wang2025,
  author  = {Wang, X. and others},
  title   = {{RadioDiff-3D}: A {3D}$\times${3D} radio map dataset and generative diffusion based benchmark for {6G} environment-aware communication},
  journal = {IEEE Trans. Netw. Sci. Eng.},
  volume  = {13},
  pages   = {3773--3789},
  month   = {Jul.},
  year    = {2025}
}

@article{diffusion,
  author  = {Fu, S. and Wu, Z. and Wu, D. and Zeng, Y.},
  title   = {Generative {CKM} construction using partially observed data with diffusion model},
  journal = {arXiv preprint arXiv:2412.14812},
  year    = {2024}
}

@article{Zhao2026,
  author  = {Zhao, L. and Wang, Y. and Wang, X. and Fei, Z. and Zeng, Y.},
  title   = {{BeamCKMDiff}: Beam-aware channel knowledge map construction via diffusion transformer},
  journal = {arXiv preprint arXiv:2601.10207},
  year    = {2026}
}

@inproceedings{DoD,
  author    = {Mo, R. and others},
  title     = {Deep machine learning-based {AoD} map and {AoA} map construction for wireless networks},
  address   = {Singapore, Singapore},
  booktitle = {IEEE Vehicular Technology Conference (VTC)},
  pages     = {1--5},
  month     = {Jun.},
  year      = {2024}
}

@article{Wu2025,
  author  = {Wu, Z. and Wu, D. and Fu, S. and Qiu, Y. and Zeng, Y.},
  title   = {{CKMImageNet}: A dataset for {AI}-based channel knowledge map toward environment-aware communication and sensing},
  journal = {IEEE Trans. Commun.},
  volume  = {73},
  number  = {12},
  pages   = {14430--14443},
  month   = {Dec.},
  year    = {2025}
}

@article{use1,
  author  = {Amiri, R. and others},
  title   = {Indoor environment learning via {RF}-mapping},
  journal = {IEEE J. Sel. Areas Commun.},
  volume  = {41},
  number  = {6},
  pages   = {1859--1872},
  month   = {Jun.},
  year    = {2023}
}

@article{PINN1,
  author  = {Karniadakis, G. E. and others},
  title   = {Physics-Informed Machine Learning},
  journal = {Nat. Rev. Phys.},
  volume  = {3},
  number  = {6},
  pages   = {422--440},
  month   = {May},
  year    = {2021}
}

@article{PINN2,
  author  = {Zhu, E. and Sun, H. and Ji, M.},
  title   = {Physics-informed generalizable wireless channel modeling with segmentation and deep learning: Fundamentals, methodologies, and challenges},
  journal = {IEEE Wireless Commun.},
  volume  = {31},
  number  = {6},
  pages   = {170--177},
  month   = {Dec.},
  year    = {2024}
}

@article{GNN1,
  author  = {Scarselli, F. and Gori, M. and Tsoi, A. C. and Hagenbuchner, M. and Monfardini, G.},
  title   = {The graph neural network model},
  journal = {IEEE Trans. Neural Netw.},
  volume  = {20},
  number  = {1},
  pages   = {61--80},
  month   = {Jan.},
  year    = {2009}
}

@inproceedings{DeepMIMO,
  author    = {Alkhateeb, A.},
  title     = {{DeepMIMO}: A generic deep learning dataset for millimeter wave and massive {MIMO} applications},
  booktitle = {Information Theory and Applications Workshop (ITA)},
  pages     = {1--8},
  month     = {Feb.},
  year      = {2019}
}

@article{openpathnet,
  title={{OpenPathNet}: An Open-Source {RF} Multipath Data Generator for {AI}-Driven Wireless Systems},
  author={Liu, L. and Chen, X. and Zhang, W.},
  journal={arXiv preprint arXiv:2512.17286},
  year={2025}
}

@inproceedings{Hamilton2017,
  author    = {Hamilton, W. and Ying, Z. and Leskovec, J.},
  title     = {Inductive representation learning on large graphs},
  booktitle = {Proc. Int. Conf. Neural Inf. Process. Syst. ({NeurIPS})},
  address   = {Long Beach, CA, USA},
  pages     = {1025--1035},
  month     = {Dec.},
  year      = {2017}
}

\end{document}